\documentclass[%
reprint,
superscriptaddress,
preprintnumbers,
nofootinbib,
nobibnotes,
amsmath,amssymb,
aps,
prd,
floatfix,
]{revtex4-2}

\usepackage{lipsum} 
\usepackage{graphicx}
\usepackage{dcolumn}
\usepackage{bm}
\usepackage{amsmath,amsfonts,amssymb,bm,mathrsfs,longtable}
\graphicspath{ {./} }
\usepackage{hyperref}
\usepackage{array}
\usepackage{color}
\usepackage[shortlabels]{enumitem}
\usepackage[utf8]{inputenc}
\usepackage[normalem]{ulem}
\usepackage{comment}
\usepackage[version=4]{mhchem}
\usepackage{outlines}
\usepackage{mathtools}
\usepackage{cleveref}
\crefname{figure}{Fig.}{Figs.}
\Crefname{figure}{Figure}{Figures}
\crefname{section}{Sec.}{Secs.}
\Crefname{section}{Section}{Sections}
\crefname{equation}{Eq.}{Eqs.}
\Crefname{equation}{Equation}{Equations}
\crefname{table}{Table}{Tables}
\hypersetup{
	colorlinks   = true,
    linkcolor={blue!70!black},
    citecolor={blue!70!black}
}

\usepackage{xcolor}
\usepackage{booktabs}
\usepackage{multirow}

\def\be{\begin{equation}}
\def\ee{\end{equation}}

\newcommand{\f}{\frac}
\newcommand{\mrm}{\mathrm}
\newcommand{\dd}{\mathrm{d}}

\newcommand*{\bigchi}{\mbox{\scalebox{1.3}{$\chi$}}}

\usepackage{aas_macros}

\begin{document}

\title{Scalar-Tensor Gravity and DESI 2024 BAO data}

\author{Angelo G. Ferrari}
 \email{anferrar@bo.infn.it}
 \affiliation{INFN, Sezione di Bologna, viale C. Berti Pichat 6/2, 40127 Bologna, Italy}
 
\author{Mario Ballardini}\email{mario.ballardini@unife.it}
 \affiliation{Dipartimento di Fisica e Scienze della Terra, Universit\`a degli Studi di Ferrara, via Giuseppe Saragat 1, 44122 Ferrara, Italy}
 \affiliation{INFN, Sezione di Ferrara, via Giuseppe Saragat 1, 44122 Ferrara, Italy}
 \affiliation{INAF/OAS Bologna, via Piero Gobetti 101, 40129 Bologna, Italy}
 
\author{Fabio Finelli}\email{fabio.finelli@inaf.it}
 \affiliation{INAF/OAS Bologna, via Piero Gobetti 101, 40129 Bologna, Italy}
 \affiliation{INFN, Sezione di Bologna, viale C. Berti Pichat 6/2, 40127 Bologna, Italy}
 
\author{Daniela Paoletti}\email{daniela.paoletti@inaf.it}
 \affiliation{INAF/OAS Bologna, via Piero Gobetti 101, 40129 Bologna, Italy}
 \affiliation{INFN, Sezione di Bologna, viale C. Berti Pichat 6/2, 40127 Bologna, Italy}

\begin{abstract}
We discuss the implications of the DESI 2024 baryon acoustic oscillations (BAO) data on scalar-tensor models of gravity. 
We consider four representative models: induced gravity (equivalent to Jordan-Brans-Dicke), where we either fix today's value of the effective gravitational constant on cosmological scales to the Newton's constant or allow them to differ, Jordan-Brans-Dicke supplemented with a Galileon term, and early modified gravity with a conformal coupling. 
In this way it is possible to investigate how different modified gravity models compare with each other when confronted with DESI 2024 BAO data.
Compared to previous analyses, for all of these models, the combination of {\em Planck} and DESI data favors a larger value of the key parameter of the theory, such as the nonminimal coupling to gravity or the Galileon term, leading also to a larger value of $H_0$, due to the known degeneracy between these parameters. 
These new results are mainly driven by the first two redshift bins of DESI.
In BDG, in which we find the largest value for $H_0$ among the models considered, 
the combination of {\em Planck} and DESI is consistent with the Chicago-Carnegie Hubble program results and reduces the $H_0$ tension with the SH0ES measurement to $1.2\sigma$ (compared to $4.5\sigma$ of $\Lambda$CDM in our {\em Planck} + DESI analysis).
\end{abstract}

\maketitle

\section{Introduction}

Baryon acoustic oscillations (BAOs) are periodic density fluctuations in the distribution of baryonic matter that originated in the early Universe.
These fluctuations were imprinted in the cosmic microwave background (CMB), as a remnant of the tightly coupled plasma of photons and baryons at high temperature.
As the Universe expanded and cooled, photons decoupled from baryons, leaving behind a fossil pattern of density ripples that act as a cosmic standard ruler on a scale of about 150 million parsecs.
The BAO signal was first detected in Sloan Digital Sky Survey (SDSS) \cite{SDSS:2005xqv}, in 2dF \cite{2dFGRS:2005yhx} galaxy surveys and has now become crucial for constraining cosmological parameters, such as the density of dark matter, the nature of dark-energy, and the geometry of the Universe.

The Dark Energy Spectroscopic Instrument (DESI) Collaboration \cite{desiurl,DESI:2016fyo} has released results from the first year BAO measurements \cite{DESI:2024uvr,DESI:2024lzq,DESI:2024mwx}, and more recently from the
full-shape galaxy clustering \citep{DESI:2024jis}. DESI 2024 BAO data include galaxies, quasars, and Lyman-$\alpha$ forest in six bins in a redshift range $0.4<z<4.1$ over a $\sim 7 500$ square degree footprint,
leading to a precision slightly better than previous Baryon Oscillation Spectroscopic Survey (BOSS) and extended BOSS (eBOSS) data \cite{DESI:2024uvr,DESI:2024lzq,DESI:2024mwx}.
The agreement with previous BOSS and eBOSS BAO data is good, except for the second and the third redshift bins of DESI \cite{DESI:2024uvr,DESI:2024lzq,DESI:2024mwx}.
This difference results in a slightly worse overall consistency with {\em Planck} DR3 data and a slightly higher value for $H_0$ inferred for the $\Lambda$ cold dark-matter ($\Lambda$CDM) model
\cite{DESI:2024uvr,DESI:2024lzq,DESI:2024mwx}.
In combination with CMB and SN data, DESI hints a preference for a phantom dark-energy parameter of state, a result with a statistical significance that has sparked vivid interest in the scientific community \cite{VanRaamsdonk:2023ion,Tada:2024znt,Yin:2024hba,Wang:2024hks,Luongo:2024fww,Cortes:2024lgw,Colgain:2024xqj,Carloni:2024zpl,Wang:2024rjd,Allali:2024cji,Giare:2024smz,Qu:2024lpx,Wang:2024dka,Yang:2024kdo,Craig:2024tky,Mukherjee:2024ryz,zabat:2024wof,Roy:2024kni,Gialamas:2024lyw,Chudaykin:2024gol,Green:2024xbb,Giare:2024gpk,Barbieri:2025moq,Chan-GyungPark:2024mlx,Chan-GyungPark:2024brx,Chan-GyungPark:2025cri,Ramadan:2024kmn,Wolf:2024eph,Wolf:2024stt,Ye:2024ywg,Bhattacharya:2024hep,Bhattacharya:2024kxp,Taule:2024bot,Colgain:2024ksa,Colgain:2024mtg}.

These results, together with inconsistencies between different measurements in the framework of the $\Lambda$CDM model \cite{Knox:2019rjx,DiValentino:2020zio,DiValentino:2020vvd,DiValentino:2021izs,Perivolaropoulos:2021jda,Schoneberg:2021qvd,Shah:2021onj,Abdalla:2022yfr}, are challenging the standard cosmological model and fueling the theoretical search for extended models.
The most relevant of these inconsistencies is the $5\sigma$ tension between the inference of $H_0$ within $\Lambda$CDM from the {\em Planck} measurement of CMB anisotropy \cite{Planck:2018nkj} and its determination with Cepheids-calibrated supernovae at low redshift \cite{Riess:2021jrx}.

Modified gravity is one of the most promising routes to reconcile these inconsistencies and possibly explain a time-varying dark-energy density.
The simplest theories of modified gravity are scalar-tensor theories (STTs), whose only additional ingredient with respect to general relativity (GR) is the presence of a scalar degree of freedom.
They were first proposed by Jordan, Brans and Dicke \cite{jordan55,Brans:1961sx}, then generalized by Horndeski who found the most general STT with second-order equations of motion \cite{Horndeski:1974wa} (for reviews see Refs.~\cite{Kase:2018aps,Kobayashi:2019hrl,Quiros:2019ktw}).
Several STTs have been studied in the context of cosmology \cite{Cooper:1981byv,Amendola:1999qq,Boisseau:2000pr,Torres:2002pe,Esposito-Farese:2000pbo,Gannouji:2006jm,Finelli:2007wb,Cerioni:2009kn}, but the relevance of the full Horndeski's theory has been appreciated only recently with its rediscovery in the context of Galileons \cite{Nicolis:2008in,Deffayet:2009wt,Deffayet:2011gz,Silva:2009km,Kobayashi:2009wr,Kobayashi:2010wa,Chow:2009fm,DeFelice:2010jn,DeFelice:2011bh}. 
The simplest STTs and more general subsets of Horndeski's theories have been analyzed with cosmological datasets by several authors \cite{Umilta:2015cta,Ballardini:2016cvy,Paoletti:2018xet,Rossi:2019lgt,Braglia:2020iik,Ballardini:2020iws,Braglia:2020auw,Ballardini:2021evv,Ballardini:2021eox,Ballardini:2023mzm,Ferrari:2023qnh,Raveri:2014cka,Bellini:2015xja,Peirone:2017vcq,SolaPeracaula:2019zsl,Benevento:2018xcu,Peirone:2019aua,Joudaki:2020shz,Kable:2021yws,Barreira:2014jha,Renk:2016olm,Frusciante:2019puu,Bessa:2023ykd,Zumalacarregui:2020cjh,HosseiniMansoori:2024pdq,Ye:2024ywg,Ye:2024kus,Wolf:2024eph,Wolf:2024stt,Bhattacharya:2024hep,Bhattacharya:2024kxp,Ramadan:2024kmn,Taule:2024bot}.

In this paper we study the impact of DESI 2024 BAO data on a selection of models within STTs of gravity.
As in previous works \cite{Umilta:2015cta,Ballardini:2016cvy,Paoletti:2018xet,Rossi:2019lgt,Braglia:2020iik,Ballardini:2020iws,Braglia:2020auw,Ballardini:2021evv,Ballardini:2021eox,Ballardini:2023mzm,Ferrari:2023qnh}, we work in the original Jordan frame, without resorting to any kind of approximations in the background or perturbations, and thus meet the accuracy required by current and future data \cite{Bellini:2014fua}, i.e. CMB and large-scale structure data which probe completely different times and scales.

The paper is organized as follows. 
In \cref{sec:model} we describe the models considered, while in \cref{sec:methods_data} we present the datasets and the details of our Markov chain Monte Carlo (MCMC) analysis.
\Cref{sec:results} is devoted to the presentation of the results.
We draw our conclusions in \cref{sec:conclusions}.

\section{A selection of scalar-tensor gravity models} \label{sec:model}
We study the following subclass of Horndeski theories \cite{Horndeski:1974wa,Deffayet:2009wt,Deffayet:2011gz,Kobayashi:2011nu,Kase:2018aps}
\be \label{eq:action}
{\cal S} = \int {\dd}^4x\sqrt{|g|} \left[G_4(\sigma)R + G_2(\sigma,X) +G_3(\sigma,X)\square\sigma + \mathcal{L_{\rm M}}\right],
\ee
where $|g|$ is the absolute value of the determinant of the metric $g_{\mu\nu}$, for which we use a $(-+++)$ signature; $\sigma$ is a time-dependent scalar field, $X \equiv -\nabla_{\mu}\sigma\nabla^{\mu}\sigma/2 = -\partial_{\mu}\sigma\partial^{\mu} \sigma/2$ is the kinetic term, and $\Box~\equiv~\nabla_\mu \nabla^\mu$ is the d'Alambert covariant operator.
$\mathcal{L_{\rm M}}$ is the matter Lagrangian which does not depend on the scalar field $\sigma$ and it is minimally coupled to the metric $g$.
The action \eqref{eq:action} predicts that gravitational waves travel at the speed of light \cite{Kase:2018aps} without any fine-tuning, in agreement with current measurements \cite{LIGOScientific:2017zic}.

We will study several subcases encompassed by~\cref{eq:action}, for which the most general form of the $G_i$ functions is
\begin{align}\label{eq:BDG_Gis}
\begin{split}
&G_4(\sigma) = F(\sigma)/2 \\
&G_3(\sigma, X) = -2 g(\sigma)X \\ 
&G_2(\sigma, X) = Z X-V(\sigma) + 4\zeta(\sigma)X^2 \,, \\
\end{split}
\end{align}
and the models we wish to discuss can be obtained with suitable choices for these functions. In the above equations $Z\equiv\pm1$ is the sign of the kinetic term.

The Einstein field equations for the theory are
\begin{align}\label{eq:EE}
    \begin{split}
        G_{\mu\nu} = \frac{1}{F(\sigma)} \bigg[ &T_{\mu\nu}^{\rm (M)} + T_{\mu\nu}^{\rm (G)} + Z \big( \partial_{\mu}\sigma\partial_{\nu}\sigma - \frac{1}{2}g_{\mu\nu}\partial^{\rho}\sigma\partial_{\rho}\sigma \big) \\ &- g_{\mu\nu} V(\sigma) + (\nabla_\mu \nabla_\nu - g_{\mu\nu} \Box) F(\sigma) \bigg] \,,
    \end{split}
\end{align}

where 
\be
T_{\mu\nu}^\mrm{(M)} = -\f{2}{\sqrt{|g|}} \frac{\delta(\sqrt{|g|}\mathcal{L}_\mrm{M})}{\delta g^{\mu\nu}}
\ee
is the energy-momentum tensor of matter fields and $T_{\mu\nu}^{(G)}$ is the energy-momentum tensor due to the Galileon term defined as 
\begin{align}
\begin{split}
T_{\mu\nu}^{\rm (G)} = -2 \Big\{ \, &g(\sigma) \, \nabla_\mu \sigma \nabla _\nu \sigma \, \Box \sigma - \nabla_{(\mu} \, \sigma \, \nabla_{\nu)} \big[\, g(\sigma) (\partial \sigma)^2\,  \big] \\ 
&+ \frac{1}{2} g_{\mu\nu} \nabla_\alpha\, \sigma \nabla^\alpha \big[\, g(\sigma) (\partial \sigma)^2 \, \big] - \frac{\zeta(\sigma)}{2} g_{\mu\nu} (\partial \sigma)^4 \\ 
&+2\zeta(\sigma) \nabla_\mu \sigma \nabla _\nu \sigma \ (\partial \sigma)^2 \, \Big\} \,,
\end{split}
\end{align}
with $\nabla_{(\mu} \, \sigma \, \nabla_{\nu)} \equiv \big(\nabla_{\mu} \, \sigma \, \nabla_{\nu} + \nabla_{\nu} \, \sigma \, \nabla_{\mu} \big)/2$. For the models with $G_3=0$ the term $T_{\mu\nu}^\mrm{(G)}$ is zero.

The equation of motion for the scalar field is obtained by varying the action~\eqref{eq:action} with respect to the scalar field $\sigma$, giving
\begin{align}\label{eq:cova_KG}
    \begin{split}
		&\Box\sigma \big[ Z - 4\zeta\, (\partial\sigma)^2 \big] -2 g\, \big\{ (\Box \sigma)^2 - \nabla^\mu\nabla^\nu \sigma \nabla_\mu\nabla_\nu \sigma \\
        &- \nabla^\mu \sigma \nabla^\nu \sigma R_{\mu\nu} \big\} +4 g,_{\sigma} \nabla^\mu \sigma \nabla^\nu \sigma \nabla_\mu \nabla_\nu \sigma 
		+ g,_{\sigma \sigma} (\partial\sigma)^4 \\
        & - 3\zeta,_{\sigma}(\partial\sigma)^4 - 4\zeta(\sigma)\nabla_\mu [ (\partial\sigma)^2 ]\nabla^\mu \sigma + \f12 F,_\sigma R -V,_\sigma = 0 \,.
    \end{split}
\end{align}

Due to the nonminimal coupling $F(\sigma)R$ in the Lagrangian, the gravitational constant in the Einstein's equations is replaced by a time-varying cosmological gravitational constant $G_\mrm{cosm} = (8\pi F)^{-1}$, which depends on the value of the scalar field $\sigma$.
It should be emphasized that the coupling constant $G_\mrm{cosm}$ is generally not the one measured between test masses. 
The effective gravitational constant $G_\mrm{eff}$, which can be measured locally, is obtained in the weak-field limit of the theory~\eqref{eq:action} and is given by \cite{Hohmann:2015kra,Quiros:2019gbt}
\be\label{eq:GeffG3}
G_\mrm{eff} = \frac{1}{16 \pi G_4} \left[ \frac{ 4 G_{4,\sigma}^2 + G_4 ( G_{2,X} -2 G_{3,\sigma} ) }{ 3 G_{4,\sigma}^2 + G_4 ( G_{2,X} -2 G_{3,\sigma})}\right] \,.
\ee
The above expression for $G_\mrm{eff}$ is valid only for theories where screening mechanism is not relevant \cite{Hohmann:2015kra}.

In this manuscript we study the following models:
\begin{enumerate}[(i)]
\item Induced gravity (IG) \cite{Cooper:1981byv,Wetterich:1987fk,Finelli:2007wb} with a quartic potential and standard kinetic term ($Z=1$) 
\begin{align}\label{eq:IG_Gis}
\begin{split}
&G_4(\sigma) = F(\sigma)/2 = \xi\sigma^2/2 \,, \\
&G_3(\sigma, X) = 0 \,, \\ 
&G_2(\sigma, X) = Z X - V(\sigma) = X - \lambda F(\sigma)^2 \,.
\end{split}
\end{align}
In this model, the effective gravitational constant between test masses~\eqref{eq:GeffG3} reduces to 
\begin{equation}
\label{eq:Geff_nmc}
G_\mrm{eff}(z=0) = \f{1}{8\pi F_0} \f{2 F_0 + 4 F^2_{0,\sigma}}{2 F_0 + 3 F^2_{0,\sigma}} \,,
\end{equation}
where the subscript $0$ means evaluated at today (at redshift $z=0$).
The value of the effective gravitational constant on small scales is determined by the cosmological evolution of the scalar field.
Therefore, we use \cref{eq:Geff_nmc} as a boundary condition for the present value of the scalar field, $\sigma_0$, thus ensuring that the effective gravitational constant between test masses in IG is consistent with the one measured in a Cavendish-type experiment: $G_\mathrm{eff}(\sigma_0) = 6.67 \times 10^{-11}\,\mathrm{m\,kg^{-1}\,s^{-2}}$.

IG is equivalent to the Jordan-Brans-Dicke (JBD) model \cite{jordan55,Brans:1961sx}. Indeed, with the field redefinition given by $\varphi\equiv\xi\sigma^2/2$ and $\xi=Z/(4\,\omega_\mathrm{BD})$, the following Horndeski functions become $G_2 = 2 \,\omega_\mrm{BD} \chi / \varphi -V(\varphi)$ and $G_4 = \varphi$, where $\chi\equiv\nabla_{\mu}\varphi\nabla^{\mu}\varphi/2$.
These are exactly the functions that define JBD gravity with the BD field $\varphi$.
\item Induced gravity + $\Delta$ ($\Delta$IG) \cite{Ballardini:2021evv} with a quartic potential. The Horndeski functions are the same as for IG but the phenomenological parameter $\Delta$ is such that 
\be\label{eq:dig_Geff}
G_\mrm{eff}=G(1+\Delta)^2 \,, 
\ee
where $G$ is the gravitational constant measured in a Cavendish-type experiment.
In this way, it is possible to allow for an imbalance between the current value of the effective gravitational constant and the measured value of the Newton’s constant on Earth. Equation~\eqref{eq:dig_Geff} corresponds to a shift of the initial or final value of the scalar field with respect to the standard IG case.
\item Brans-Dicke Galileon (BDG) \cite{Ferrari:2023qnh} in the phantom branch given by the $G_i$'s of \cref{eq:BDG_Gis} with the choice $Z=-1$, meaning that the kinetic term is not canonical. Nevertheless, this model is not affected by ghost instabilities \cite{Ferrari:2023qnh}. 
For concreteness, the functions of the field $F(\sigma), V(\sigma), g(\sigma), \zeta(\sigma)$ are, respectively
\begin{align}
&F(\sigma)=\xi\sigma^2 \,, \\
&V(\sigma)=\Lambda \,, \\ 
&g(\sigma)=\alpha \sigma^{-1} \,, \\
&\zeta(\sigma)=\alpha \sigma^{-2} = g(\sigma)\sigma^{-1} \,.
\end{align}
Thus, the explicit form of the Horndeski functions is the following:
\begin{align}\label{eq:BDG_Gis_explicit}
\begin{split}
&G_4(\sigma) = \xi\sigma^2/2 \,, \\
&G_3(\sigma, X) = -2 g(\sigma)X = -\frac{2\,\alpha}{\sigma} X \,, \\ 
&G_2(\sigma, X) = - X - \Lambda + \frac{4\,\alpha}{\sigma^{2}} X^2 \,. \\ 
\end{split}
\end{align}
The relation between $g(\sigma)$ and $\zeta(\sigma)$ is such that by using the BD field $\varphi\equiv\xi\sigma^2/2$ with $\xi=Z/(4\omega_\mathrm{BD})$ one obtains the following Horndeski functions: $G_2 = 2 \omega_\mrm{BD} \chi / \varphi -V(\varphi)$, $G_3 = -2 g(\varphi)\chi$, and $G_4 = \varphi$, hence the name Brans-Dicke Galileon.
Due to the Vainshtein screening mechanism \cite{Vainshtein:1972sx}, the theory reduces to GR on small scales in the sense that the post-Newtonian parameters are those of GR and the gravitational constant in screened regions is $G=G_{\rm cosm}$ \cite{Kimura:2011dc}. 
The cosmological evolution of the scalar field determines the value of the coupling constant of GR in screened regions. 
We ensure that $G_{\rm cosm}(\sigma_0)$ is equal to the gravitational constant measured today on small scales $G$, thus restoring GR with the correct value of the gravitational constant on small scales today.
\item Early modified gravity with conformal coupling (EMG-CC) \cite{Braglia:2020auw}, characterized by a coupling $F(\sigma)$ and a generalized potential $V_\mrm{EMG}(\sigma)$, given by 
\begin{align}
\begin{split}
&F(\sigma)=M_\mrm{Pl}^2+\xi\sigma^2 \,, \\  
&V_\mrm{EMG}(\sigma) = \Lambda + \lambda\sigma^4/4 \,,
\end{split}
\end{align}
where $\lambda \equiv 10^{2V_0}/M_{\rm Pl}^4 $.
For simplicity we consider the case of conformal coupling $\xi=-1/6$ with $Z=+1$.
The explicit form of the Horndeski function is thus
\begin{align}\label{eq:EMG_Gis_explicit}
\begin{split}
&G_4(\sigma) = \left(M_\mrm{Pl}^2 - \sigma^2/6\right)/2 \,,\\
&G_3(\sigma, X) = 0 \,,\\ 
&G_2(\sigma, X) = X - \Lambda - \lambda\sigma^4/4 \,. \\ 
\end{split}
\end{align}
In EMG the scalar field decays to the minimum of the potential, i.e. $\sigma = 0$.
Thus, $G_\mathrm{eff}(z=0) = G$ and the post-Newtonian parameters reduce to those of GR without any fine-tuning.
Consequently, the second free parameter of the theory is the initial value of the field $\sigma_\mathrm{ini}$.
\end{enumerate}

Since the cosmological effects of the theories just described have been extensively studied in the past \cite{Umilta:2015cta, Ballardini:2016cvy,Paoletti:2018xet,Rossi:2019lgt,Braglia:2020iik,Ballardini:2020iws,Braglia:2020auw,Ballardini:2021evv}, we do not discuss their dynamics in detail and do not present the equations for the linear perturbations, which can be found in the works cited above; see for instance Refs.~\cite{Ballardini:2023mzm,Ferrari:2023qnh}. Nevertheless, we present the equations of motion in a spatially flat Friedmann-Lema\^itre-Robertson-Walker (FLRW) background in Appendix~\ref{sec:flrw_equations}.

\section{Methodology and datasets}\label{sec:methods_data}

In order to derive the constraints on the cosmological parameters we perform a MCMC analysis by using the publicly available sampling code \texttt{Cobaya} \cite{Torrado:2020dgo,2019ascl.soft10019T}, connected to our extension of the code \texttt{CLASSig} \cite{Umilta:2015cta} (an Einstein-Boltzmann solver based on \texttt{CLASS} \cite{2011arXiv1104.2932L,Blas:2011rf}).
For the sampling, we use the Metropolis-Hastings algorithm with a Gelman-Rubin \cite{Gelman:1992zz} convergence criterion $R-1 < 0.01$.
The reported means values and uncertainties of the parameters, together with the contour plots have been obtained using \texttt{GetDist} \cite{getdisturl,Lewis:2019xzd}.

We make use of CMB anisotropy data in temperature, polarization, and lensing from {\em Planck} data release $3$ \cite{Planck:2019nip,Planck:2018lbu}.
We consider the {\em Planck} baseline likelihood (hereafter P18), which consists of the {\tt Plik} likelihood at high multipoles, $\ell>30$, the {\tt commander} likelihood at lower multipoles for temperature and {\tt SimAll} for the E-mode polarization \cite{Planck:2019nip}, and the conservative multipole range, $8< L < 400$, for the CMB lensing. 

\begin{figure}
\includegraphics[width=0.5\textwidth]{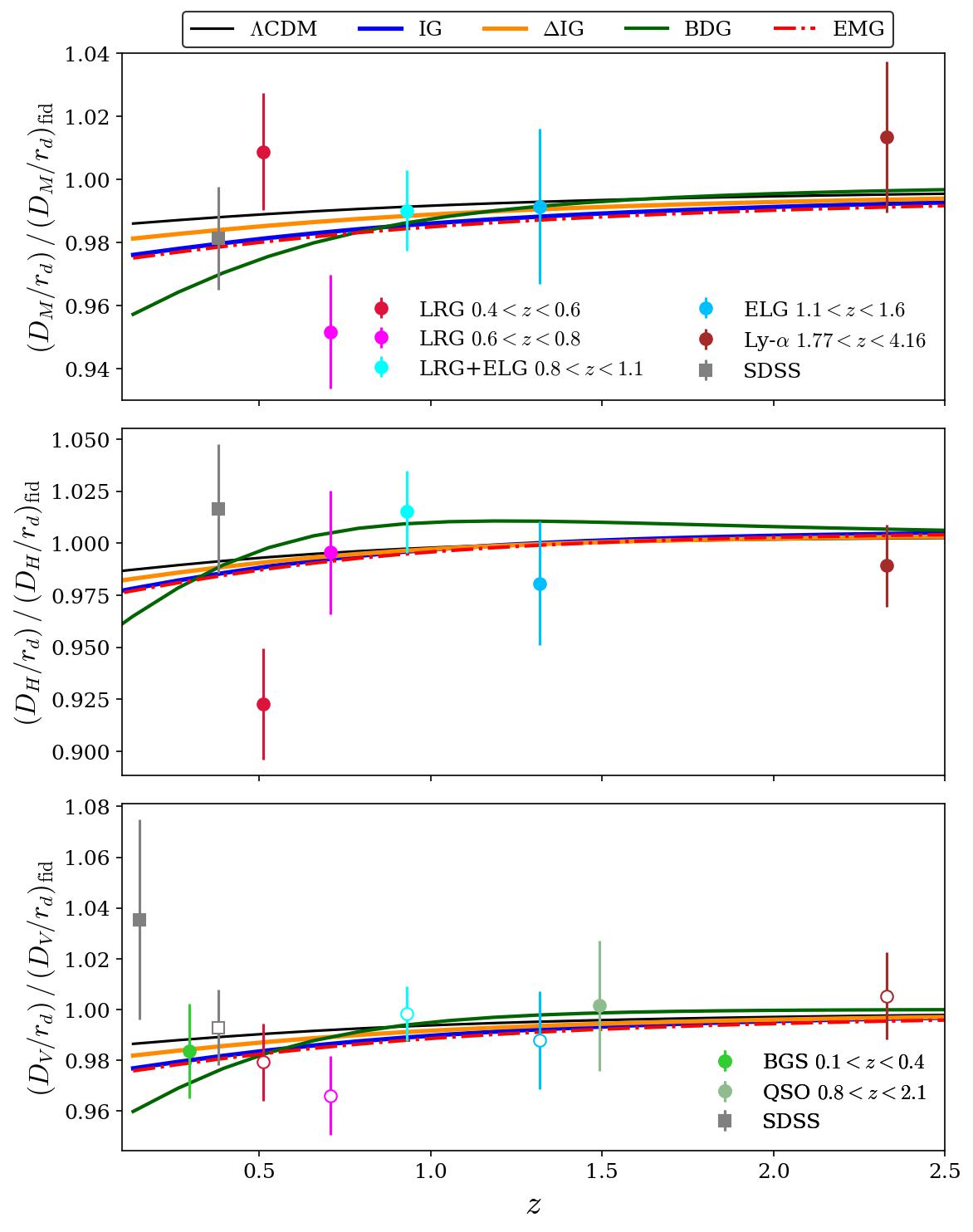}
\caption{BAO distance scales measured by the DESI Collaboration and the SDSS points at low redshift we use in the analysis. These points are the SDSS DR7 main galaxy sample measurement at $z_\mathrm{eff}=0.15$ \cite{Ross:2014qpa} and the BOSS galaxy sample results at $z_\mathrm{eff}=0.38$ \cite{BOSS:2016apd,BOSS:2016sne,BOSS:2016hvq}. 
From top to bottom, the panels show $D_M/r_d, D_H/r_d$ and $D_V/r_d$, all relative to the respective quantities evaluated in the DESI fiducial cosmology described in \cite{DESI:2024uvr}. 
For some redshift bins of DESI and SDSS, only $D_V/r_d$ measurements were possible and they are shown in the third panel.
Open markers in the third panel indicate repetition of information already contained in the top two panels.}
\label{fig:bao_points_and_bestfits}
\end{figure}

We use the latest BAO dataset from DESI \cite{DESI:2024uvr,DESI:2024lzq,DESI:2024mwx} (see \cref{fig:bao_points_and_bestfits}). 
It consists of BAO distance scales measured in several redshift bins using different tracers: the bright galaxy sample (BGS) at $0.1<z<0.4$, the luminous red galaxy sample (LRG) at $0.4<z<0.6$ and $0.6<z<0.8$, a combined sample of LRG and emission line galaxies (ELG) at $0.8<z<1.1$, the ELG sample at $1.1<z<1.6$, the quasar sample (QSO) from redshift $z=0.8$ to $z=2.1$, and finally the Lyman-$\alpha$ forest sample (Ly$\alpha$) at redshift $1.77<z<4.16$. We refer to this dataset simply as DESI.
In addition, as done in Ref.~\cite{DESI:2024mwx}, we consider the composite dataset by combining SDSS results at low redshifts ($z<0.6$)~\cite{Ross:2014qpa,BOSS:2016apd,BOSS:2016sne,BOSS:2016hvq,eBOSS:2020yzd} and DESI measurements at $z>0.6$.
This means that the BGS sample and the first bin of LRG from DESI are excluded from the analysis. 
Following Ref.~\cite{DESI:2024mwx}, we refer to this dataset as DESI+SDSS. 
Where possible, the {\em Planck} plus DESI and {\em Planck} plus (DESI+SDSS) results are compared with previous ones obtained by {\em Planck} in combination with previous compilations of BAO data, which include results from 6dF \cite{2011MNRAS.416.3017B}, SDSS DR7 \cite{Ross:2014qpa}, SDSS DR12 \cite{BOSS:2016apd,BOSS:2016sne,BOSS:2016hvq,eBOSS:2020yzd}, which covered $0.8 \lesssim z \lesssim 1.1$ as redshift range, and eBOSS DR14 \cite{deSainteAgathe:2019voe,Blomqvist:2019rah}.
We note that these are all postreconstruction BAO data. It would be interesting to investigate whether the results remain stable with prereconstruction BAO or with different estimation techniques, such as the Linear Point \cite{Anselmi:2015dha,Anselmi:2017cuq}.

We also use the combination with Gaussian priors given by the supernovae and $H_0$ for the dark-energy equation of etate (SH0ES) program \cite{Riess:2021jrx}: ${H_0 = \left(73.04 \pm 1.04\right)\, \rm km\,s^{-1}\,Mpc^{-1}}$, and by the Chicago-Carnegie Hubble Program (CCHP) \cite{Freedman:2024eph}: $H_0 = \left[69.96 \pm 1.05 \text{ (stat)}\, \pm 1.12 \text{ (sys)}\right]\, \rm km\,s^{-1}\,Mpc^{-1}$. 
Using a prior directly on $H_0$ rather than on the peak absolute magnitude of type Ia supernovae $M_\mathrm{B}$ is equivalent for these classes of models, as demonstrated in Ref.~\cite{Ballardini:2021eox}.

Given the datasets just described, the combinations we have considered are
\begin{enumerate}[(i)]
    \item P18 + DESI,
    \item P18 + (DESI+SDSS),
    \item P18 + DESI + SH0ES,
    \item P18 + DESI + CCHP.
\end{enumerate}

We sample on six standard parameters: $\Omega_{\rm b} h^2$, $\Omega_{\rm c} h^2$, $H_0$, $\tau_{\rm reio}$, $\ln\left(10^{10} A_{\rm s}\right)$, and $n_{\rm s}$, and the modified gravity parameters, in particular
\begin{enumerate}[(i)]
    \item for IG we sample on $\zeta_{\rm IG} \equiv \ln(1+ 4\xi)$ in the prior range $[0, 0.039]$. The $\Lambda$CDM limit for IG is recovered for $\xi\rightarrow0$.
    \item For $\Delta$IG we sample on $\zeta_{\rm IG}$ in the prior range $[0, 0.039]$ and on $\Delta \in [-0.3, 0.3]$. The $\Lambda$CDM limit for $\Delta$IG is at $\xi\rightarrow0$ and $\Delta=0$.
    \item For BDG we keep the nonminimal coupling parameter fixed to $\xi=5\times10^{-5}$ as in Ref.~\cite{Ferrari:2023qnh}.
    This value chosen to be large enough that it would not respect the Solar System constraints on the post-Newtonian parameters in absence of $G_3$ and screening.
    We allow only for the amplitude of the Galileon term to vary. 
    To this purpose we define $1/\widetilde\alpha_8\equiv10^{-8}\widetilde\alpha$, where $\widetilde\alpha = \alpha \times {\rm (Mpc\, [GeV]}^{-1} )^{-2}$.
    We sample on $1/\widetilde\alpha_8$ because the $\Lambda\rm CDM$ limit is obtained when $\widetilde\alpha \rightarrow \infty$, corresponding to $1/\widetilde\alpha \rightarrow 0$ \cite{Ferrari:2023qnh}.
    In this way, our prior range $1/\widetilde\alpha_8 \in [0, 0.4]$ includes the $\Lambda$CDM limit at a finite value. 
    \item For EMG-CC, the additional parameters on which we sample are $\sigma_\mathrm{ini} / M_\mathrm{Pl} \in [0, 0.9]$ and $V_0 \in [-4, 3.5]$.
\end{enumerate}
In addition to the cosmological parameters, we sample on the nuisance and foreground parameters of the P18 likelihood.
In our analysis, we assume two massless neutrinos with $N_{\rm eff} = 2.0328$ and one massive neutrino with fixed mass $m_\nu = 0.06\, {\rm eV}$.
As in Ref.~\cite{Ballardini:2016cvy}, we fix the primordial \ce{^{4}He} mass fraction $Y_{\rm p}$ taking into account the different value of the effective gravitational constant during big bang nucleosynthesis (BBN) in the presence of a nonminimally coupled scalar field, and the baryon fraction $\omega_{\rm b}$, tabulated in the public code
{\tt PArthENoPE} \cite{Pisanti:2007hk,Consiglio:2017pot}. 

Moreover, for each run we compute the best-fit values, obtained minimizing the $\chi^2$ with the \texttt{BOBYQA} method \cite{Powell2009TheBA,cartis2019improving,cartis2022escaping} implemented in \texttt{Cobaya}. We quote the difference in the model $\chi^2$ with respect to $\Lambda \rm CDM$, i.e. $\Delta\chi^2 \equiv \chi^2-\chi^2_{\Lambda\rm CDM}$. Thus, negative values of $\Delta\chi^2$ indicate an improvement in the fit with respect to $\Lambda\rm CDM$. We also compute the Akaike information criterion (AIC) \cite{1100705} of the extended model $\mathcal M$ relative to that of $\Lambda$CDM: 
$\Delta {\rm AIC} = \Delta \chi^2 + 2 (N_{\mathcal M} - N_{\Lambda\rm CDM})$, where $N_{\mathcal M}$ is the number of free parameters of the model. The number of additional parameters, i.e. $N_{\mathcal M} - N_{\Lambda\rm CDM}$, is one for IG and BDG and it is two for $\Delta$IG and EMG-CC.

\section{Results}\label{sec:results}
\begin{figure}
\includegraphics[width=0.47\textwidth]{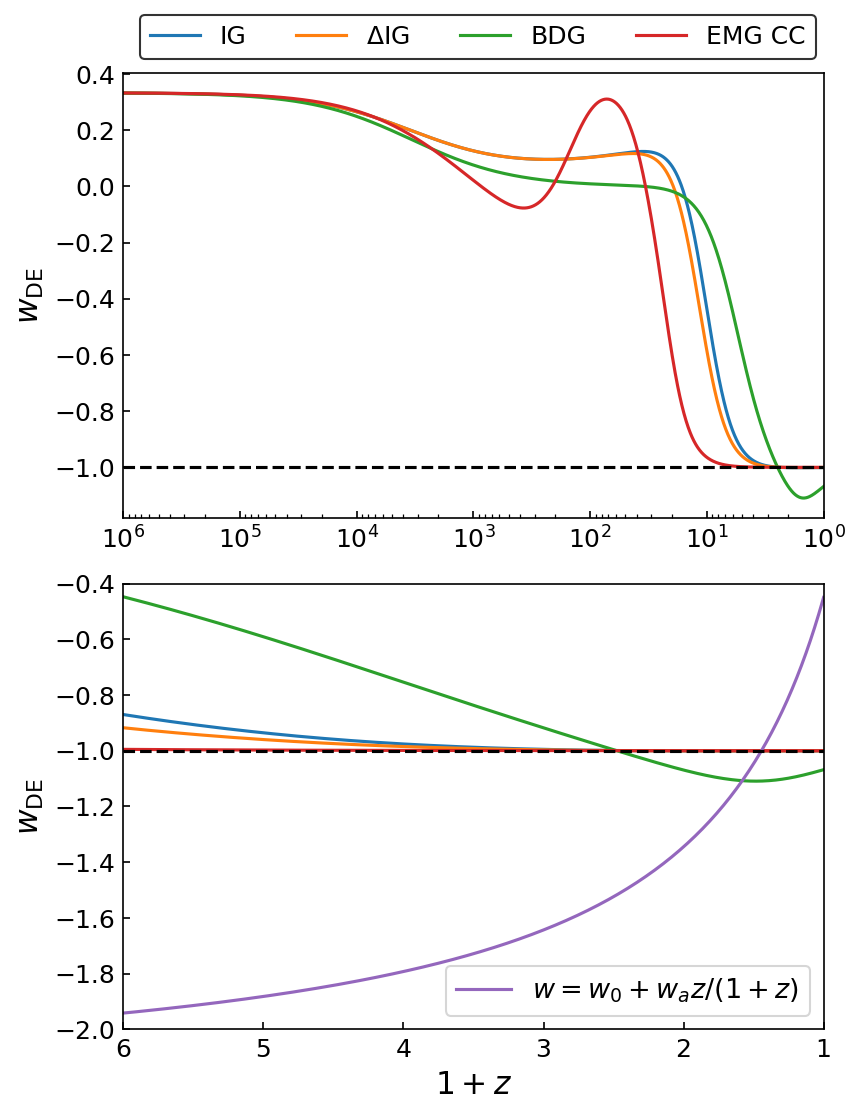}
\caption{The equation of state parameter for dark-energy computed in a mimicking Einstein gravity \cite{Boisseau:2000pr} in all the models at the best-fit values obtained with the P18+DESI dataset. The bottom panel is a low-redshift enlargement of the top panel.
In the bottom panel, we also plot the equation of state parameter $w=w_0 + w_a z/(1+z)$, with $w_0 = -0.45$ and $w_a = -1.79$. These are the mean values of the 1D marginalized constraints of the $w_0w_a$CDM model obtained by the DESI Collaboration in their P18+DESI analysis \cite{DESI:2024mwx}. 
Since the definitions of $w$ and $w_\mathrm{DE}$ as given in Appendix~\ref{sec:flrw_equations} differ at high redshift, they can only be compared at low $z$.
}
\label{fig:wde_all_models}
\end{figure}

This section examines the constraints on the parameters of the scalar-tensor models introduced in~\cref{sec:model} and discusses their implications for the cosmological tensions and the overall fit to the observational data, based on the datasets detailed in \cref{sec:methods_data}.

A key finding for all models is a preference for larger values of the extra parameter of the theory -- either the nonminimal coupling to gravity or the Galileon term -- when combining DESI data with {\em Planck}, compared to previous analyses based on the SDSS BAO dataset.
This preference is due to a degeneracy between these parameters and the Hubble constant leading to an increase in the inferred value of $H_0$.
In addition, the DESI data show a preference for a phantom dark-energy equation of state \cite{DESI:2024mwx}. 

In~\cref{fig:wde_all_models}, we present the evolution of the equation-of-state for dark-energy as defined in~\cref{eq:wde}, computed for all the models at the best-fit values for the P18+DESI dataset.
The parameter $w_{\rm DE}$ follows the dominant component of the Universe: deep in the radiation epoch it has a value close to $1/3$ , then in the matter era it decreases toward zero, and finally in the present epoch it becomes negative, $w_{\rm DE} \simeq -1$, mimicking a cosmological constant.
The bump present in the EMG-CC model is due to an increasing value of the kinetic term of the scalar field, consistent with the fact that the field oscillates strongly at these redshifts.
As it can be seen in the figure, only BDG can generate a phantom equation of state and consequently it emerges as the model that best-fits the combined datasets.
In the bottom panel, the purple line represents the equation of state parameter $w=w_0 + w_a z/(1+z)$, with $w_0 = -0.45$ and $w_a = -1.79$. These are the mean values of the 1D marginalized constraints of the $w_0w_a$CDM model obtained by the DESI Collaboration in their P18+DESI analysis \cite{DESI:2024mwx}
As shown by the purple line, the phantom behavior is quite different in BDG and $w_0w_a$CDM.
It is more pronounced in the $w_0w_a$CDM case and it also crosses the phantom divide to reach values $>-1$ today.

Tables and plots with the results presented in the next sections can be found respectively in Appendix~\ref{sec:tables} and Appendix~\ref{sec:plots}.

\subsection{IG and $\Delta$IG}\label{sec:ig_results}
\begin{figure*}
\includegraphics[width=\textwidth]{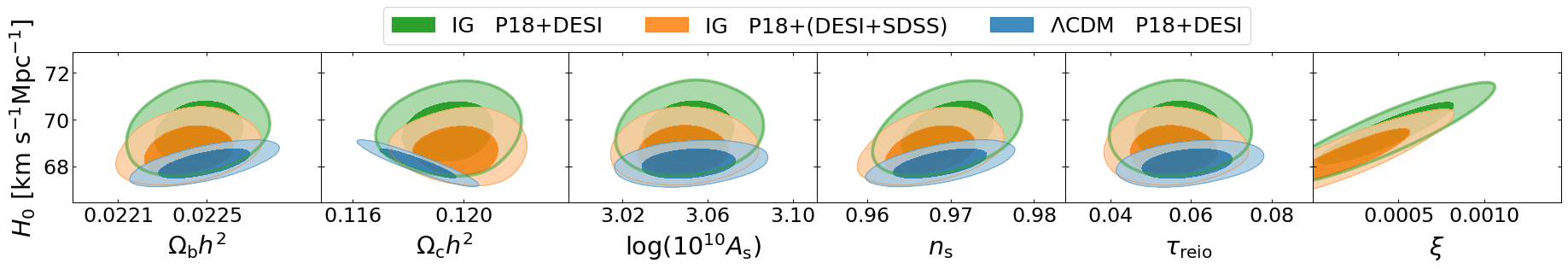}
\caption{Marginalized $68\%$ and $95\%$ 2D credible regions comparing P18+DESI and P18+(DESI+SDSS) datasets for the IG and $\Lambda$CDM models.}
\label{fig:rect_H0_ig_desi_sdss}
\end{figure*}
\begin{figure*}
\includegraphics[width=\textwidth]{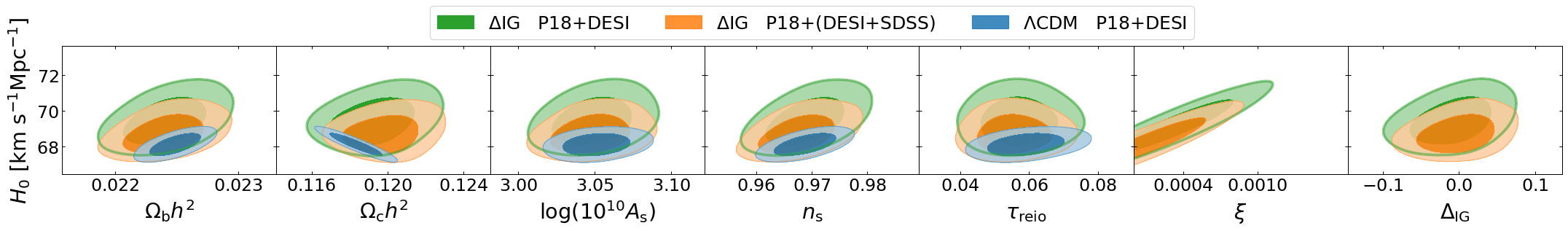}
\caption{Marginalized $68\%$ and $95\%$ 2D credible regions comparing P18+DESI and P18+(DESI+SDSS) datasets for the $\Delta$IG and $\Lambda$CDM models.}
\label{fig:rect_H0_dig_desi_sdss}
\end{figure*}
This subsection discusses the constraints on the parameters of the induced gravity and induced gravity with the imbalance $\Delta$, obtained using the CMB and BAO datasets. 
Both models have a quartic potential, with the $\Delta$IG model extending IG by introducing the phenomenological parameter $\Delta$ discussed in~\cref{sec:model} and in Ref.~\cite{Ballardini:2021evv}.

In IG, the parameter encapsulating deviations from GR is constrained by P18+DESI to be $\xi < 0.00088$ (95\% credible interval, CI). This is not the tightest bound obtained by using cosmological datasets. 
In fact, in Ref.~\cite{Ballardini:2020iws} $\xi < 0.00055$ was found with a combination of {\em Planck} CMB data and BAO from the SDSS data release 12 and earlier ones.
The reason for this is not the lack of constraining power of DESI, but its preference for a larger value of $H_0$ relative to the SDSS. 
The Hubble constant is in turn correlated with $\xi$ (see the right panel of~\cref{fig:rect_H0_ig_desi_sdss}), and the net result is that the bound on the nonminimal coupling parameter $\xi$ is weaker with this dataset.
In particular, for the Hubble constant we find that the 68\% CI is $H_0 = 69.52^{+0.79}_{-0.90}\,{\rm km\, s^{-1} Mpc^{-1}}$, indicating a reduced tension with the SH0ES measurement to $2.6\sigma$. 

With the combination P18+(DESI+SDSS), which is constructed by replacing the first two redshift bins of DESI with SDSS low-$z$ results, the mean and marginalized uncertainty on $H_0\,\left[{\rm km\, s^{-1} Mpc^{-1}}\right]$ becomes $68.75^{+0.61}_{-0.76}$, which is reflected in a tighter 95\% limit on the nonminimal coupling parameter: $\xi<0.00067$. 

Consistently with what has just been discussed, the constraints on the derived parameters depending on $\xi$ are weaker
in this work compared to Ref.~\cite{Ballardini:2020iws}.
We report the results for the deviation from unity of the first post-Newtonian parameter: the 95\% CI is $1-\gamma_\mathrm{PN} < 0.0035$ for P18+DESI and ${< 0.0027}$ for P18+(DESI+SDSS). The 95\% CI on the variation of the cosmological gravitational constant between the radiation era and the present time is
\begin{align}
\begin{split}
\frac{\delta G_\mathrm{cosm}}{G_\mathrm{cosm}}(z=0) > -0.026\quad &\text{ P18+DESI} \,, \\
\frac{\delta G_\mathrm{cosm}}{G_\mathrm{cosm}}(z=0) > -0.020\quad &\text{ P18+(DESI+SDSS)} \,.
\end{split}
\end{align}
The limits on the time derivative of the gravitational constant at the present time and on the other cosmological parameters are reported in~\cref{tab:ig_table} for IG.

Statistical comparison shows no preference for IG over $\Lambda$CDM with the CMB + BAO datasets used here.

The inclusion of the phenomenological parameter $\Delta$ allows for a broader exploration of deviations from GR: see~\cref{fig:rect_H0_dig_desi_sdss}.
The results for this parameter are consistent with zero with the 68\% CI corresponding to $\Delta=-0.008\pm 0.035$ (P18+DESI) and $\Delta=-0.002\pm 0.033$ [P18+(DESI+SDSS)].
For the coupling parameter $\xi$, we obtain, with P18+DESI $\xi <0.00093$ (95\% CI) and $< 0.00073$ (95\% CI) with P18+(DESI+SDSS).
The second bound obtained by combining DESI and SDSS is tighter than the one reported in Ref.~\cite{Ballardini:2021eox} for a P18+BAO dataset.
The marginalized mean and uncertainty for the Hubble constant with P18+DESI is $69.47^{+0.76}_{-0.97}\,\mathrm{km\, s^{-1}\, Mpc^{-1}}$, which reduces the tension with the SH0ES measurement to $2.6\sigma$.
As with IG, the inclusion of SDSS data in our analysis, shifts the value of $H_0\,\left[{\rm km\, s^{-1} Mpc^{-1}}\right]$ at a slightly lower value: $68.74^{+0.61}_{-0.80}$, in $3.5\sigma$ tension with the SH0ES measurement. 
In~\cref{tab:dig_table}, in addition to the post-Newtonian parameter $\gamma_\mathrm{PN}$, the variation of the gravitational constant and its time derivative at the present time, we also quote the ratio between the effective gravitational constant $G_\mathrm{eff}$ and $G$, since in this model they differ as the consistency condition on $\sigma_0$ is not respected; see~\cref{eq:dig_Geff}.

We do not observe a preference for $\Delta$IG over $\Lambda$CDM for the combinations of P18 with the two BAO datasets considered.

\subsection{BDG}\label{sec:bdg_results}
\begin{figure*}
\includegraphics[width=\textwidth]{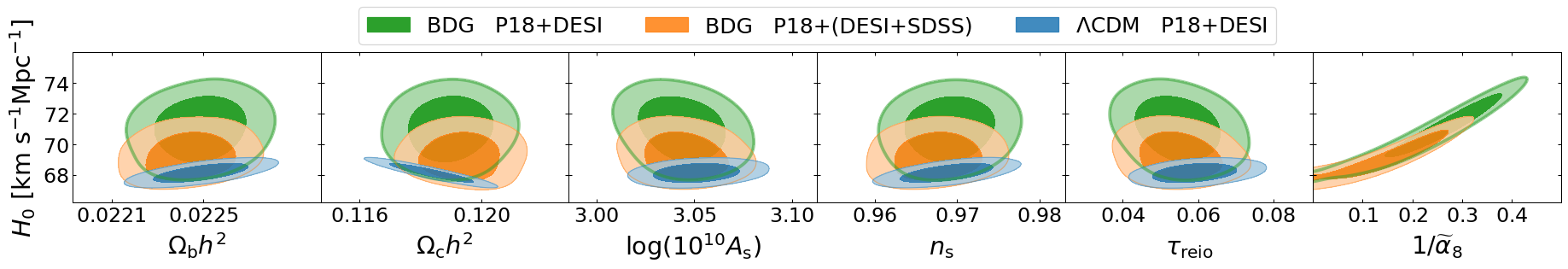}
\caption{Marginalized $68\%$ and $95\%$ 2D credible regions comparing P18+DESI and P18+(DESI+SDSS) datasets for the BDG and $\Lambda$CDM models.}
\label{fig:rect_H0_alpha_BDG_desi_sdss}
\end{figure*}
We present in this section the results for the BDG model.
\Cref{fig:rect_H0_alpha_BDG_desi_sdss} shows the marginalized 2D posterior distributions for BDG compared with $\Lambda$CDM, in the $p$ -- $H_0$ planes, where $p$ is one of the cosmological parameters we fitted to the data. 
The posterior distributions for most of the cosmological parameters are close to the $\Lambda$CDM values.
As shown in~\cref{fig:rect_H0_alpha_BDG_desi_sdss}, the only parameter that differs significantly from its $\Lambda$CDM value is the Hubble constant, due to the shape of the posterior in the $1/\widetilde\alpha_8$ -- $H_0$ plane.
This behavior was already shown in Ref.~\cite{Ferrari:2023qnh} with a different BAO dataset.
Due to the degeneracy between $1/\widetilde\alpha_8$ and $H_0$, the latter can take significantly larger values with respect to the inference in $\Lambda$CDM. 
In particular, for the combination P18+DESI, the marginalized $68\%$ CI is 
\be\label{eq:H0_bdg_p18_desi}
H_0 = 71.0^{+1.5}_{-1.3}\;{\rm km\, s^{-1}\, Mpc^{-1}}, 
\ee
indicating a significant reduction of the Hubble tension: the value is consistent at $1.2\sigma$ with the SH0ES local measurement. 
The first DESI results \cite{DESI:2024mwx} already indicated that this dataset allows a larger $H_0$ within $\Lambda$CDM.
In BDG, the additional parameter $1/\widetilde\alpha_8$ further pushes the $H_0$ constraint upward, as the data prefer nonzero values for $1/\widetilde\alpha_8$, deviating from its $\Lambda$CDM limit. 
In particular, we find that the $68\%$ CI is $1/\widetilde\alpha_8 = 0.255^{+0.095}_{-0.064}$, which is about $3\sigma$ away from the $\Lambda$CDM value of 0.
The previous determination of the Galileon parameter from CMB and BAO measured by SDSS was only an upper limit: $1/\widetilde\alpha_8 < 0.28$ (95\% CI) \cite{Ferrari:2023qnh}.
With DESI BAO we are able to constrain the parameter by a factor of $2$ better.
These values of $1/\widetilde\alpha_8$ can produce a phantom equation of state for dark-energy ($w_\mathrm{DE}<-1)$, as shown in \cref{fig:wde_all_models}, which does not compromise the fit to DESI, showing the preference of the DESI BAO for dynamical (and phantom) dark-energy.

\begin{figure}
\includegraphics[width=0.45\textwidth]{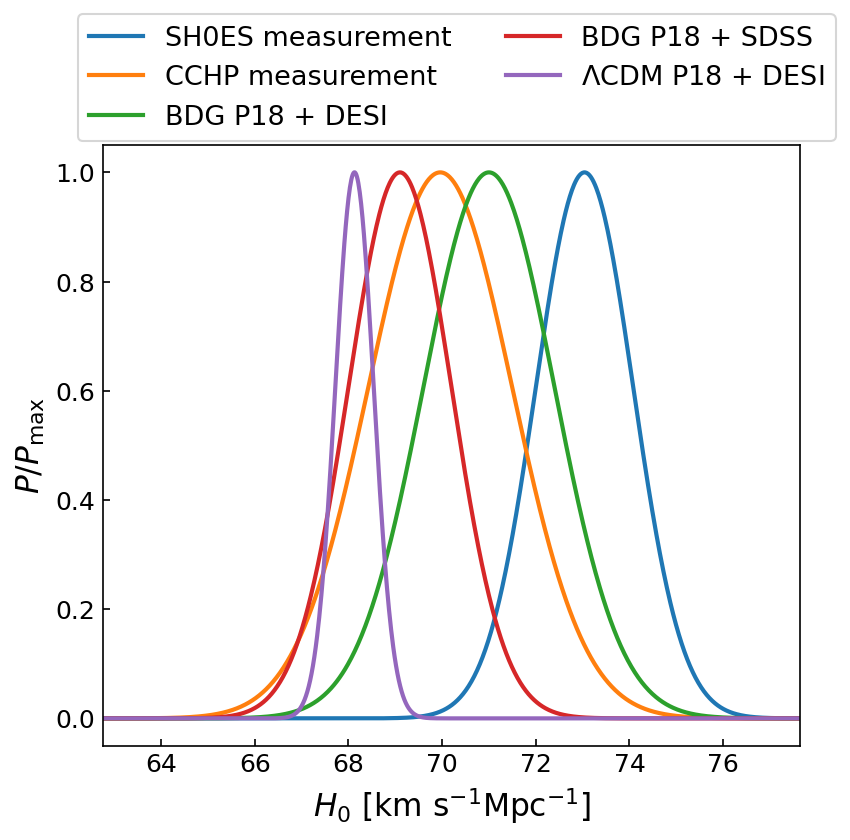}
\caption{SH0ES and CCHP local measurements of the Hubble constant together with the marginalized 1D posterior for $H_0$ in BDG and $\Lambda$CDM, using the dataset P18+DESI. For BDG it also presented the result obtained in Ref.~\cite{Ferrari:2023qnh} using P18+SDSS. The SDSS dataset was composed of the postreconstruction measurements from BOSS DR12 \cite{BOSS:2016wmc}, low-$z$ BAO measurements from the SDSS DR7 main galaxy sample (MGS) and 6dF  \cite{Ross:2014qpa,2011MNRAS.416.3017B}, Ly$\alpha$ BAO measurements from eBOSS DR14, and combination of those \cite{deSainteAgathe:2019voe,Blomqvist:2019rah,Cuceu:2019for}.}
\label{fig:bdg_p18_desi_H0_1d_vs_shoes_cchp}
\end{figure}

Considering the combination P18+(DESI+SDSS), which replaces the data from first two redshift bins of DESI with SDSS data at low redshift, the analysis is qualitatively similar. The Hubble constant is moved toward larger values due to the pull of $1/\widetilde\alpha_8$, but the quantitative results are less significant.
We find \mbox{$H_0\, $ = $69.2^{+0.9}_{-1.2}\,{\rm km\, s^{-1}\, Mpc^{-1}}$}, and $1/\widetilde\alpha_8=0.157^{+0.088}_{-0.076}$, consistent with 0 at $2\sigma$.
Additionally, the determination of $1/\widetilde\alpha_8$ is tighter than the one of Ref.~\cite{Ferrari:2023qnh}.
We can quantify the shift toward larger values of the Hubble constant by using the Hubble tension (with respect to SH0ES) as a proxy: in the P18+DESI case the tension is at the $1.2\sigma$ level, effectively eliminating any significant disagreement with local measurements (see also \cref{fig:bdg_p18_desi_H0_1d_vs_shoes_cchp}).
Instead, for P18+(DESI+SDSS) we find a $2.6\sigma$ tension. This still highlights ability of the model to accommodate larger $H_0$ values with respect to $\Lambda$CDM.
On the other hand, it leaves some tension with local measurements.
\Cref{fig:bdg_p18_desi_H0_1d_vs_shoes_cchp} shows the local measurements of the Hubble constant by the SH0ES and CCHP groups, together with the marginalized 1D posterior for $H_0$ in BDG, using the dataset P18+DESI. 
Our result lies right in the middle of the two local measurements and it is consistent with both.

We now discuss the statistical fit of BDG compared to the standard cosmological model. 
For P18 + DESI we have $\Delta\chi=-5.6$ and $\Delta \rm AIC = {-3.6}$, showing a mild preference with respect to $\Lambda$CDM.
This improvement is driven by the CMB with $\Delta\chi^2\mathrm{(P18)} = {-5.8}$, but it would not be possible without the inclusion of DESI BAO data.
In fact, CMB alone is not able to constrain $1/\alpha_8$, and for this reason, larger values are allowed, if not preferred.
When combining CMB with DESI BAO data, these large values are still allowed without compromising the fit to DESI $\left[\Delta\chi^2\mathrm{(DESI)} = {0.2}\right]$.
On the contrary, when combining (DESI+SDSS), these large values of $1/\widetilde\alpha_8$ are not allowed by BAO measurements.
Therefore, for the CMB we have $\Delta\chi^2\mathrm{(P18)} = {-2.7}$ (compared to $-5.8$ discussed above), while for BAO we obtain $\Delta\chi^2\mathrm{(DESI+SDSS)} = {0.3}$.
The total is $\Delta\chi=-2.4$, which shows no preference with respect to $\Lambda$CDM.

As discussed in depth in Refs.~\cite{Tsujikawa:2019pih, Zumalacarregui:2020cjh}, coupled Galileons can be challenged by observations of the local Universe, such as lunar laser ranging (LLR) experiments.
In these models, the time derivative of the scalar field may be nonzero today, and since $\dot G_\mathrm{cosm} \propto -\dot\sigma$, the time derivative of the gravitational constant can be larger than LLR observations \cite{Hofmann:2018myc} : $\dot G / G = (7.1\pm7.6)\times 10^{-14}\,\mathrm{yr}^{-1}$.
Assuming a homogeneous evolution of the local time derivative of the scalar field, we obtain the following 68\% CI: $\dot{G}_\mathrm{cosm}/G_\mathrm{cosm} (z=0) = \left(\,-11.1^{+2.4}_{-3.9}\,\right)\times 10^{-13}$ for P18 + DESI and $\dot{G}_\mathrm{cosm}/G_\mathrm{cosm} (z=0) = \left(\,-7.1^{+3.1}_{-3.9}\,\right)\times 10^{-13}$ for P18+(DESI+SDSS), which is consistent with $0$ at the $2\sigma$ level.
These values exceed the LLR limits.
However, it should be noted that the LLR constraint is correlated with the vector describing the rotation of the Moon's core \cite{Hofmann:2018myc}, which is poorly constrained independently of $\dot G$. 
The LLR limit assumes a core rotation vector obtained using $\dot G = 0$.
An analysis without this assumption could lead to weaker LLR constraints on the time derivative of the gravitational constant.
Additionally, the results obtained in our analysis assume a homogeneous evolution of the local field derivative; if the local evolution is inhomogeneous, the time variation of Newton's constant could be locally very suppressed \cite{Zumalacarregui:2020cjh}.

The results on all the other cosmological parameters can be found in \cref{tab:bdgph_table}.

\subsection{EMG-CC}\label{sec:emg_results}
\begin{figure*}
\includegraphics[width=\textwidth]{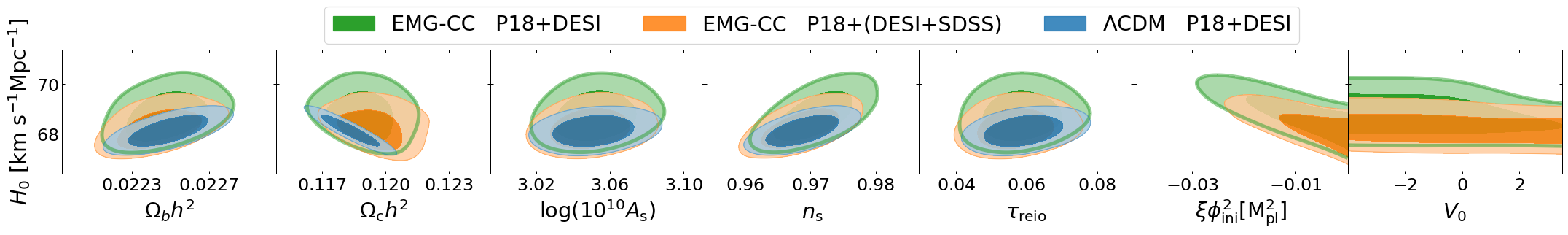}
\caption{Marginalized $68\%$ and $95\%$ 2D credible regions comparing P18+DESI and P18+(DESI+SDSS) datasets for the EMG-CC and $\Lambda$CDM models.}
\label{fig:rect_H0_emgcc_desi_sdss}
\end{figure*}

Using the combination of CMB (P18) and BAO datasets, the 95\% CI for initial value of the scalar field in EMG-CC is constrained to be 
\begin{align}
\sigma_\mathrm{ini} [M_\mathrm{Pl}] < 0.38 \quad &\text{P18+DESI} \,, \\
\sigma_\mathrm{ini} [M_\mathrm{Pl}] < 0.34 \quad &\text{P18+(DESI+SDSS)} \,.
\end{align}
The amplitude of the potential, $V_0$, is instead unconstrained, as can also be seen from \cref{fig:rect_H0_emgcc_desi_sdss}, where we present the 2D marginalized posterior distribution in the $p$ -- $H_0$ planes ($p$ is any of the cosmological or modified gravity parameters fitted to the data).
As the figure shows, for a given dataset, slightly higher values of $H_0$ are preferred with respect to the $\Lambda$CDM: for P18 + DESI we obtain that the $65\%$ CI for the Hubble constant is $H_0 = 68.74^{+0.58}_{-0.72}\, \rm km\,s^{-1}\,Mpc^{-1}$, which alleviates the tension to $3.5\sigma$.
This is the lowest value of $H_0$ obtained with P18+DESI among the models we analyzed, making it less suitable for solving the Hubble tension.
This cannot be said in general for EMG, since for simplicity we considered here the model with the restriction of conformal coupling ($\xi=-1/6)$.
In fact, without this restriction, EMG and more generally nonminimally coupled models with $\xi>0$ (not considered here) are able to significantly reduce the Hubble tension \cite{Braglia:2020iik,Braglia:2020auw,FrancoAbellan:2023gec}.

EMG-CC is not statistically preferred by the combination of CMB and BAO with respect to $\Lambda$CDM: we find, for P18 + DESI $\Delta\chi^2=-2.0\, (\Delta\text{AIC}=2.0)$ and for P18 + (DESI+SDSS) $\Delta\chi^2=-1.9\, (\Delta\text{AIC}=2.1)$.

As noted in \cref{sec:model}, in EMG the post-Newtonian parameters and the effective gravitational constant on small scales are those of GR.
For this reason we do not quote them in \cref{tab:emg_cc_table}, where the results for all the parameters and combinations of datasets are presented.

\subsection{Priors on $H_0$ from local measurements}\label{sec:H0_prior}
This section examines the impact of imposing a Gaussian prior on $H_0$ on our cosmological results, as discussed in \cref{sec:methods_data}. 
The inferred value of the $H_0$ in BDG using P18+DESI is consistent with both the SH0ES and the CCHP local measurements, as shown in \cref{fig:bdg_p18_desi_H0_1d_vs_shoes_cchp}.
Hence, we first focus on BDG in the present section and then quote the results obtained for the other models. 
\subsubsection{BDG}
\begin{figure}
\includegraphics[width=0.4\textwidth]{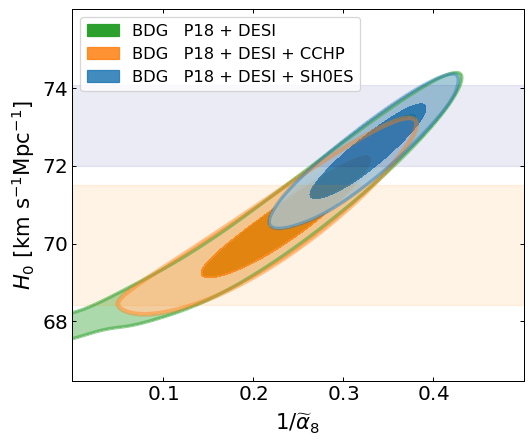}
\caption{Marginalized 68\% and 95\% 2D credible regions for BDG in the $H_0$--$1/\widetilde\alpha_8$ plane for the datasets P18+DESI, P18+DESI+SH0ES, and P18+DESI+CCHP. The shaded bands represent the $1\sigma$ local measurements of SH0ES (blue) and CCHP (orange).}
\label{fig:bdg_p18_desi_H0_priors_H0-alpha8_plane}
\end{figure}
In~\cref{fig:bdg_p18_desi_H0_priors_H0-alpha8_plane} we present the results of the analysis for the BDG model: the addition of the Gaussian prior pulls the credible interval of $1/\widetilde\alpha_8$ away from zero, indicating a stronger preference for BDG with respect to $\Lambda$CDM.
With the SH0ES prior, both an upward shift in the mean of $H_0$ and a reduction in the error bars contribute to this effect.
The CCHP prior also reduces the error bars, but does not significantly shift the mean of $H_0$.
Nevertheless, the tighter constraints result in a 2D contour that is not consistent with $1/\widetilde\alpha_8 = 0$.

The values of the Hubble constant for BDG that we obtain, in units of ${\rm km\, s^{-1}\, Mpc^{-1}}$, are

\begin{align}
H_0 & = 70.6\pm 1.0 \quad &&\text{P18 + DESI + CCHP} \,, \label{eq:h_p18_desi_cchp}\\
H_0 & = 72.33\pm 0.80 &&\text{P18 + DESI + SH0ES} \,, \label{eq:h_p18_desi_shoes}
\end{align}
see~\cref{fig:bdg_p18_desi_H0_priors_H0-alpha8_plane}.

Note that the value reported in \cref{eq:h_p18_desi_cchp} is lower than the results \eqref{eq:H0_bdg_p18_desi} of the P18+DESI analysis. 
This is expected, since the CCHP prior itself is smaller than the P18+DESI inference (\cref{fig:bdg_p18_desi_H0_1d_vs_shoes_cchp}), and it pulls the Hubble constant toward its lower central value.
The results in \cref{eq:h_p18_desi_cchp,eq:h_p18_desi_shoes} are consistent with the corresponding local measurements from which we have imposed a prior, but they are also consistent with each other at the $1.1\sigma$ level.
In particular, if we quantify the Hubble tension with respect to the SH0ES measurement, the result in \cref{eq:h_p18_desi_cchp} is only in $1.7\sigma$ tension with it.
Thus, the addition of the CCHP prior increases the Hubble tension (relative to SH0ES) from $1.2\sigma$ (for P18+DESI) to $1.7\sigma$, which is still very consistent with the SH0ES measurement.

When priors coming from local measurements are included, the BDG model is preferred to $\Lambda$CDM with increased statistical significance.
The improvement in goodness of fit is mild for P18+DESI+CCHP, for which we find an overall $\Delta\chi=-8$, mostly driven by CMB data.
For P18+DESI\allowbreak+SH0ES we instead find $\Delta\chi=-23.4$, driven by both CMB $\left[\Delta\chi^2\mathrm{(P18)} = -11.4\right]$ and by the previous $\left[\Delta\chi^2\mathrm{(SH0ES)}=-15.3\right]$.
The fit to DESI BAO worsens: $\Delta\chi^2\mathrm{(DESI)} = 3.8$, compared to $0.2$ in the P18+DESI analysis. 
However, this degradation is outweighed by the improvements observed in the other datasets. 
As discussed in \cref{sec:bdg_results}, the CMB data allow large values of $1/\widetilde\alpha_8$, which are correlated with higher values of $H_0$. 
Given this correlation, the inclusion of the SH0ES prior further increases the $1/\widetilde\alpha_8$, moderately improving the CMB fit and the overall $\chi^2$, despite the deterioration in DESI BAO. 
Consequently, the ability of the BDG model to accommodate a larger Hubble constant ultimately leads to the overall improvement in the $\chi^2$.
With one additional parameter in BDG compared to $\Lambda$CDM, we obtain $\Delta \rm{AIC} = -21.4$, showing a preference with respect to the standard model of cosmology.

\subsubsection{IG and $\Delta$IG}
We now discuss the effect of the Gaussian prior in IG, $\Delta$IG and EMG-CC.
The trends are qualitatively similar: the detection of the modified gravity parameters is more stringent due to the degeneracies with $H_0$, highlighted in \cref{sec:ig_results,sec:emg_results}.
This is also reflected in a statistical preference for all models with respect to $\Lambda$CDM when the SH0ES prior is considered.

In IG, P18+DESI+SH0ES gives $H_0=(71.00\pm0.69)\,{\rm km\, s^{-1}\, Mpc^{-1}}$, while the 68\% CI for $\xi$ is $0.00085\pm 0.00022$, showing a detection of $\xi\neq0$, as already observed in Ref.~\cite{Ballardini:2021eox}.
While these results are more constraining than those of Ref.~\cite{Ballardini:2020iws}, they cannot be compared because they use as prior the previous $H_0 = (74.03 \pm 1.42)\, \rm km\, s^{-1} Mpc^{-1}$ of Ref.~\cite{Riess:2019cxk}.
With the inclusion of the SH0ES prior, the fit to the standard cosmological model improves significantly: $\Delta\chi^2=-15.2$ and $\Delta \rm AIC=-13.2$.
Using the CCHP prior (P18+DESI+CCHP), $H_0=(69.59\pm0.73)\, \rm km\, s^{-1} Mpc^{-1}$, and $\xi$ is constrained to be $0.00048^{+0.00022}_{-0.00025}$.

If the parameter $\Delta$ is allowed to vary in IG, it is consistent with zero in all cases: the 68\% CI corresponds to $\Delta=0.019^{+0.029}_{-0.033}$ (P18+DESI+SH0ES) and $\Delta=-0.008\pm 0.035$ (P18+DESI+CCHP).
For the coupling parameter $\xi$, with P18+DESI+SH0ES $\xi=0.00092\pm 0.00025$ and with P18+DESI+CCHP $\xi < 0.00089$ (95\% CI).
With SH0ES and CCHP added to P18+DESI, we get for $H_0\,[{\rm km\, s^{-1}\, Mpc^{-1}}]$, $71.14\pm 0.75$ and $69.56^{+0.70}_{-0.84}$.
As discussed for other models, when the SH0ES prior is included, we obtain a substantially negative $\Delta\chi^2$ of $-17.4$, with $\Delta\rm AIC = -13.4$, again showing a preference over $\Lambda$CDM.

\subsubsection{EMG-CC}
In EMG-CC, with the addition of the SH0ES prior to P18 and DESI, there is a sharper detection of the initial value of the scalar field: its 68\% CI is $\sigma_\mathrm{ini} [M_\mathrm{Pl}] = 0.38^{+0.08}_{-0.06}$.
This is due to the degeneracy between the nonminimal coupling function (and hence $\sigma_\mathrm{ini}$) and $H_0$.
Since larger values of the Hubble constant are favored by the SH0ES prior, the constraint on the initial value of the scalar field is shifted along the degeneracy, leading to the detection of $\sigma_\mathrm{ini}\neq0$. 
With this combination of datasets it is also possible to impose a a 95\% credible upper bound on the amplitude of the potential, $V_0<1.2$.
This is the only case where we can constrain the potential in EMG-CC. 
Using the SH0ES prior, the 68\% CI for the Hubble constant is $H_0 = (70.27\pm 0.65)\, \rm km\, s^{-1}\, Mpc^{-1}$.
As can be seen from \cref{tab:emg_cc_table}, our results for the combination P18+DESI+CCHP are very similar to those obtained for P18+DESI, showing the consistency of these datasets, when used to constrain EMG-CC.
The constraint on $H_0\; [\rm km\, s^{-1}\, Mpc^{-1}]$ for P18+DESI+CCHP is $69.0^{+1.3}_{-1.2}$.

\section{Conclusions}\label{sec:conclusions}
We have studied the implications of DESI 2024 BAO data in combination with {\em Planck} to constrain a selection of scalar-tensor models such as induced gravity (IG, equivalent to Jordan-Brans-Dicke), Brans-Dicke Galileon (BDG), and early modified gravity with conformal coupling (EMG-CC).
We have explored how modifications to gravity due to these scalar-tensor models can affect cosmological parameters, in particular the Hubble constant, and resolve tensions between datasets.

One of the key findings of this paper is that the combination of {\em Planck} and DESI data leads to larger values of the additional parameters of the theory (nonminimal coupling to gravity or the Galileon term) compared to previous analyses which employed pre-DESI BAO data.
This preference leads to an increase in the inferred value of Hubble constant compared to previous studies for all the models we analyzed.

In particular, the results can be summarized as follows:
\paragraph{IG and $\Delta$IG} In both the constrained and unconstrained cases, the combination of Planck and DESI relaxes the constraint on the coupling parameter $\xi$ with respect to previous studies, due to a preference of DESI data for larger $H_0$ and the degeneracy between $H_0$ and $\xi$.
This results in a mild alleviation of the Hubble tension but no statistical preference over $\Lambda$CDM.
Furthermore, we do not observe any deviation of $\Delta$ from zero.
\paragraph{BDG} The combination P18+DESI tightens the constraint on the Galileon parameter, $1/\widetilde\alpha_8 = 0.255^{+0.095}_{-0.064}$, away from its $\Lambda$CDM null value (only an upper bound could be found in \cite{Ferrari:2023qnh}). 
As a consequence the Hubble constant is increased to $H_0 = 71.0^{+1.5}_{-1.3}\, \mathrm{km\, s^{-1}\, Mpc^{-1}}$. Therefore, the BDG model significantly reduces the tension in $H_0$ with respect to the SH0ES measurement, agreeing with it at the $1.2\sigma$ level.
The model also predicts a phantom-like dark-energy equation of state ($w<-1$), consistent with DESI’s preference for dynamical dark-energy, and shows a mild statistical preference over $\Lambda$CDM ($\Delta\mathrm{AIC}=-3.6$).
\paragraph{EMG-CC} The constraints on EMG-CC remain less restrictive: for P18+DESI we find that the amplitude of the potential $V_0$ is unconstrained, and an upper bound on the initial value of the scalar field whose 95\% CI is $\sigma_\mathrm{ini} [M_\mathrm{Pl}] < 0.38$.
Since we restrict to $\xi=-1/6$, the Hubble tension is not significantly relaxed in this model, in agreement with previous works.
Concerning the fit to the data, we do not observe any statistical preference for $\Lambda$CDM.

\vspace{3ex}
We have also analyzed the models by using the combination of DESI and SDSS data, as done in Ref.~\cite{DESI:2024mwx}, by replacing the lowest redshift bins of the DESI dataset with results from SDSS \cite{eBOSS:2020yzd}.
This combination of datasets reduces the tendency for larger values of the modified gravity parameters and $H_0$, but it does not completely reverse it.
For example, the parameter $1/\widetilde\alpha_8$ is still allowed to reach larger values with respect to the analysis of Ref.~\cite{Ferrari:2023qnh}, and as a consequence the inference of $H_0$ is only in tension with the SH0ES measurement at the $1.7\sigma$ level.

Since the $H_0$ tension is relaxed in all the models considered, we have also included Gaussian priors from local measurements of $H_0$ in our analysis.
We focused mainly on the BDG model since the results obtained by using P18+DESI are consistent with both the SH0ES and CCHP local measurement.
In this model, by adding the SH0ES prior we obtain, for the Hubble constant in $[\mathrm{km\, s^{-1} Mpc^{-1}}]$, $H_0 = 72.33\pm 0.80$, while with the CCHP prior we get $H_0 = 70.6\pm 1.0$.
These results are in agreement with each other and with the local measurements used to impose the prior.
For the combination P18+DESI+SH0ES, we observe a stronger departure from the $\Lambda$CDM limit of the theories in all models: $\xi = 0.00085\pm 0.00022$ in IG, $\xi = 0.00092\pm 0.00025 \text{ and } \Delta = 0.019^{+0.029}_{-0.033}$ in $\Delta$IG, $1/\widetilde\alpha_8 = 0.326\pm 0.042$ in BDG and $\sigma_\mathrm{ini} [M_\mathrm{Pl}] = 0.38^{+0.08}_{-0.06}$ in EMG-CC.
The inclusion of the SH0ES measurement statistically favors these scalar-tensor models with respect to $\Lambda$CDM: $\Delta\chi^2=-15.2$ for IG, $\Delta\chi^2=-17.4$ for $\Delta$IG, $\Delta\chi^2=-23.4$, for BDG, and $\Delta\chi^2=-$ for EMG-CC.

A broader exploration of the parameter space of these theories, such as the possibility of simultaneously varying the coupling to the Ricci scalar $\xi$ and the Galileon term in BDG, is left for a future work, as the study of the neutrino sector in these scalar-tensor models. 
These future analyses will also benefit from the incorporation of DESI full-shape clustering measurements \cite{DESI:2024jis} and upcoming data from several observational campaigns such as {\em Euclid}, the Vera Rubin Observatory, and the Simons Observatory; see Refs.~\cite{Alonso:2016suf,Ballardini:2019tho}. 
These future data will play a crucial role in refining the constraints on dark-energy and probing a redshift evolution of the Newton's constant, as predicted by scalar-tensor modifications of gravity.

\section*{Acknowledgments}
M.B., F.F. and D.P. acknowledge financial support from the ASI/INAF Contract for the Euclid mission n.2018-23-HH.0, from the INFN InDark initiative and from the COSMOS network through the ASI (Italian Space Agency) Grants No. 2016-24-H.0 and No. 2016-24-H.1-2018, as well as No. 2020-9-HH.0 (participation in LiteBIRD phase A). 

This work has made use of computational resources of CNAF HPC cluster in Bologna.

\clearpage
\appendix
\onecolumngrid
\section{Tables}\label{sec:tables}
In this section we present \cref{tab:ig_table,tab:dig_table,tab:bdgph_table,tab:emg_cc_table} with the results on the cosmological parameters obtained in our analysis.
\begin{table*}[h!]
\caption{Constraints on the main parameters ($68\%$ CI unless otherwise stated) for IG with $V(\sigma) \propto F(\sigma)^2$ considering the combinations: P18 + DESI, P18 + (DESI+SDSS), P18 + DESI + SH0ES, and P18 + DESI + CCHP.}
\setlength{\tabcolsep}{6pt}
{\footnotesize
\begin{ruledtabular}
\begin{tabular} { l  c c c c}
\multicolumn{5}{c}{IG}\\
\midrule
{} & P18 + DESI & P18 + (DESI+SDSS) & P18 + DESI + SH0ES  & P18 + DESI + CCHP\\
\midrule
{\boldmath$\ln(10^{10} A_\mathrm{s})$}        & $3.051\pm 0.014       $ & $3.049\pm 0.014       $ & $3.054\pm 0.014    $ & $3.051\pm 0.014$                      \\
{\boldmath$n_\mathrm{s}$}                      & $0.9697\pm 0.0037     $ & $0.9677\pm 0.0035     $ & $0.9719\pm 0.0035  $ & $0.9699\pm 0.0036$                    \\
{\boldmath$H_0 [\mathrm{km\, s^{-1} Mpc^{-1}}]$} & $69.52^{+0.79}_{-0.90}$ & $68.75^{+0.61}_{-0.76}$ & $71.00\pm 0.69     $ & $69.59\pm 0.73$                       \\
{\boldmath$\Omega_\mathrm{b} h^2$}             & $0.02246\pm 0.00013   $ & $0.02242\pm 0.00013   $ & $0.02250\pm 0.00013$ & $0.02247\pm 0.00013$                  \\
{\boldmath$\Omega_\mathrm{c} h^2$}             & $0.1195\pm 0.0011     $ & $0.1198\pm 0.0010     $ & $0.1198\pm 0.0011  $ & $0.1195\pm 0.0011$                    \\
{\boldmath$\tau_\mathrm{reio}$}                & $0.0570\pm 0.0072     $ & $0.0557\pm 0.0072     $ & $0.0570\pm 0.0074  $ & $0.0570\pm 0.0073$                    \\
{\boldmath$\zeta_\mathrm{IG}$}                 & $< 0.0035$ (95\%)   & $< 0.0027$ (95\%)   & $0.0034\pm 0.0009$ & $0.0019^{+0.0016}_{-0.0018}$ (95\%)\\
\midrule
$S_8$   & $0.818\pm 0.010$ & $0.824\pm 0.010$ & $0.815\pm 0.010$ & $0.818\pm 0.010$\\
{$\xi$} & $< 0.00088$ (95\%) & $< 0.00067$ (95\%) & $0.00085\pm 0.00022$ & $0.00048^{+0.00022}_{-0.00025}$\\

$\delta G_\mathrm{cosm}/G_\mathrm{cosm}(z=0)$ & $> -0.026$ (95\%) & $> -0.020$ (95\%) & $-0.0248\pm 0.0063$ & $-0.014^{+0.013}_{-0.012}$ (95\%)\\

$10^{13}{\dot{G}}_\mathrm{cosm}/G_\mathrm{cosm}(z=0) [\rm yr^{-1}]$ & $> -1.07$ (95\%) & ${> -0.82}$ (95\%) & ${-1.02\pm 0.26}$ & ${-0.58^{+0.30}_{-0.26}}$\\

$1-\gamma_\mathrm{PN}$ & $< 0.0035$ (95\%) & ${< 0.0027}$ (95\%) & ${0.0034\pm 0.0009}$ & $0.0019^{+0.0016}_{-0.0018}$ (95\%)\\

\midrule
$\Delta\bigchi^2\mathrm{(P18)}$ & ${-0.1}$ & $-1.5$ & $-5.2$  & $-1.3$ \\ 
$\Delta\bigchi^2\mathrm{(BAO)}$ & ${-1.8}$ & $-0.7$ & $0.7$   & $-1.3$ \\ 
$\Delta\bigchi^2(H_0)$          & {---}    & ---    & $-10.6$ & $-1.2$ \\ 
$\Delta\bigchi^2$               & ${-1.8}$ & $-2.2$ & $-15.2$ & $-3.8$ \\ 
$\Delta \rm AIC $               & ${0.2} $ & $-0.2$ & $-13.2$ & $-1.8$ \\ 
\end{tabular}
\end{ruledtabular}
}
\label{tab:ig_table}
\end{table*}

\begin{table*}[h!]
\caption{Constraints on the main parameters ($68\%$ CI unless otherwise stated) for $\Delta$IG with $V(\sigma) \propto F(\sigma)^2$ considering the combinations: P18 + DESI, P18 + (DESI+SDSS), P18 + DESI + SH0ES, and P18 + DESI + CCHP.}
\setlength{\tabcolsep}{6pt}
{\footnotesize
\begin{ruledtabular}
\begin{tabular} { l  c c c c}
\multicolumn{5}{c}{$\Delta$IG}\\
\midrule
{} & P18 + DESI & P18 + (DESI+SDSS) & P18 + DESI + SH0ES  & P18 + DESI + CCHP\\
\midrule
{\boldmath$\ln(10^{10} A_\mathrm{s})$}        & $3.049\pm 0.017$             & $3.048\pm 0.017$             &  $3.059\pm 0.017$                & $3.049\pm 0.017$        \\  
{\boldmath$n_\mathrm{s}$}                      & $0.9691\pm 0.0047$           & $0.9676\pm 0.0045$           &  $0.9735^{+0.0044}_{-0.0050}$    & $0.9694\pm 0.0045$      \\
{\boldmath$H_0 [\mathrm{km\, s^{-1} Mpc^{-1}}]$} & $69.47^{+0.76}_{-0.97}$      & $68.74^{+0.61}_{-0.80}$      &  $71.14\pm 0.75$                 & $69.56^{+0.70}_{-0.84}$ \\
{\boldmath$\Omega_\mathrm{b} h^2$}             & $0.02242\pm 0.00022$         & $0.02241\pm 0.00021$         &  $0.02259^{+0.00021}_{-0.00023}$ & $0.02243\pm 0.00021$    \\
{\boldmath$\Omega_\mathrm{c} h^2$}             & $0.1193^{+0.0014}_{-0.0015}$ & $0.1198^{+0.0012}_{-0.0014}$ &  $0.1204\pm 0.0016$              & $0.1193\pm 0.0015$      \\
{\boldmath$\tau_\mathrm{reio}$}                & $0.0569\pm 0.0074$           & $0.0556\pm 0.0071$           &  $0.0569\pm 0.0074$              & $0.0570\pm 0.0074$      \\
{\boldmath$\zeta_\mathrm{IG}$}                 & $< 0.0037$ (95\%)            & $< 0.0029$ (95\%)            &  ${0.0037\pm 0.0010}$            & ${<0.0036}$ (95\%)      \\
{\boldmath$\Delta$}                            & $-0.008\pm 0.035$            & $-0.002\pm 0.033$            &  $0.019^{+0.029}_{-0.033}$       & $-0.008\pm 0.035$       \\
\midrule
$S_8$   & $0.819\pm 0.011$  & $0.824\pm 0.010$   & $0.814\pm 0.010$      & $0.818\pm 0.010$  \\
{$\xi$} & $<0.00093$ (95\%) & $< 0.00073$ (95\%) &  $0.00092\pm 0.00025$ & $< 0.00089$ (95\%)\\

$\delta G_\mathrm{cosm}/G_\mathrm{cosm}(z=0)$ & ${> -0.027}$ (95\%) & ${> -0.021}$ (95\%) & ${-0.027\pm 0.007}$ & ${> -0.026}$ (95\%)\\

$10^{13}{\dot{G}}_\mathrm{cosm}/G_\mathrm{cosm}(z=0) [\rm yr^{-1}]$ & ${> -1.12}$ (95\%) & ${> -0.88}$ (95\%) & ${-1.10\pm 0.31}$ & ${> -1.08}$ (95\%)\\

$G_\mathrm{eff}/G(z=0)$ & $0.985\pm 0.070$ & $0.998\pm 0.067$ & $1.039^{+0.058}_{-0.068}$  & $0.986\pm 0.069$\\

$1-\gamma_\mathrm{PN}$ & ${< 0.00367}$ (95\%) & ${< 0.0029}$ (95\%) & ${0.0036\pm 0.0010}$  & ${<0.0035}$ (95\%)\\

\midrule
$\Delta\bigchi^2\mathrm{(P18)}$ & $-2.5$ & $-1.9$ & $-5.2 $ & $-2.1$ \\ 
$\Delta\bigchi^2\mathrm{(BAO)}$ & $-1.2$ & $-0.8$ & $1.8  $ & $-1.3$ \\ 
$\Delta\bigchi^2(H_0)$          & ---    & ---    & $-14.0$ & $-1.1$ \\ 
$\Delta\bigchi^2$               & $-3.7$ & $-2.7$ & $-17.4$ & $-4.6$ \\ 
$\Delta \rm AIC $               & $0.3 $ & $1.3 $ & $-13.4$ & $-2.6$ \\ 
\end{tabular}
\end{ruledtabular}
}
\label{tab:dig_table}
\end{table*}

\begin{table*}[htb]
\caption{Constraints on the main parameters ($68\%$ CI unless otherwise stated) for BDG phantom ($Z=-1$) and $g(\sigma) = \alpha\sigma^{-1}$ considering the combinations: P18 + DESI, P18 + (DESI+SDSS), P18 + DESI + SH0ES, and P18 + DESI + CCHP.}
\setlength{\tabcolsep}{8pt}
\begin{ruledtabular}
\begin{tabular} { l  c c c c}
\multicolumn{5}{c}{BDG}\\
\midrule
{} & P18 + DESI & P18 + (DESI+SDSS) & P18 + DESI + SH0ES  & P18 + DESI + CCHP\\
\midrule
{\boldmath$\ln(10^{10} A_\mathrm{s})$}        & $3.044\pm 0.015$          & $3.045\pm 0.014$          & $3.040\pm 0.015$     & $3.045\pm 0.015$         \\
{\boldmath$n_\mathrm{s}$}                      & $0.9691\pm 0.0036$        & $0.9675\pm 0.0036$        & $0.9694\pm 0.0036$   & $0.9690\pm 0.0036$       \\
{\boldmath$H_0 [\mathrm{km\, s^{-1} Mpc^{-1}}]$} & $71.0^{+1.5}_{-1.3}$      & $69.2^{+0.9}_{-1.2}$    & $72.33\pm 0.80$      & $70.6\pm 1.0$            \\
{\boldmath$\Omega_\mathrm{b} h^2$}             & $0.02249\pm 0.00013$      & $0.02244\pm 0.00013$      & $0.02251\pm 0.00013$ & $0.02248\pm 0.00013$     \\
{\boldmath$\Omega_\mathrm{c} h^2$}             & $0.1190\pm 0.0009$      & $0.1194\pm 0.0009$      & $0.1190\pm 0.0009$ & $0.1190\pm 0.0009$     \\
{\boldmath$\tau_\mathrm{reio}$}                & $0.0547\pm 0.0075$        & $0.0549\pm 0.0074$        & $0.0529\pm 0.0074$   & $0.0552\pm 0.0076$       \\
{\boldmath$1/\widetilde\alpha_8$}              & $0.255^{+0.095}_{-0.064}$ & $0.157^{+0.088}_{-0.076}$ & $0.326\pm 0.042$     & $0.231^{+0.072}_{-0.055}$\\
\midrule
$S_8$ &$0.812\pm 0.010$ & $0.819\pm 0.010$ & $0.809\pm 0.010$ & $0.813\pm 0.010$\\

$\delta G_\mathrm{cosm}/G_\mathrm{cosm}(z=0)$ & $-0.010^{+0.002}_{-0.003}$ & $-0.006^{+0.003}_{-0.003}$ & $-0.012^{+0.002}_{-0.002}$ & $-0.009^{+0.002}_{-0.003}$\\

$10^{13}{\dot{G}}_\mathrm{cosm}/G_\mathrm{cosm}(z=0) [\rm yr^{-1}]$ & $-11.1^{+2.4}_{-3.9}$ & $-7.1^{+3.1}_{-3.9}$ & ${-13.9^{+1.5}_{-1.8}}$ & $-10.2^{+2.2}_{-2.9}$\\
\midrule
$\Delta\bigchi^2\mathrm{(P18)}$ & ${-5.8}$ & $-2.7$ & $-11.4$ & $-6.3$ \\ 
$\Delta\bigchi^2\mathrm{(BAO)}$ & ${0.2}$  & $0.3$  & $3.2$   & $0.1$  \\ 
$\Delta\bigchi^2(H_0)$ & {---}             & ---    & $-15.3$ & $-0.8$ \\ 
$\Delta\bigchi^2$ & ${-5.6}$               & $-2.4$ & $-23.4$ & $-7.0$ \\ 
$\Delta \rm AIC $ & ${-3.6}$               & $-0.4$ & $-21.4$ & $-5.0$ \\ 
\end{tabular}
\end{ruledtabular}
\label{tab:bdgph_table}
\end{table*}

\begin{table*}[htb]
\caption{Constraints on the main parameters ($68\%$ CI unless otherwise stated) for EMG-CC ($\xi=-1/6$) considering the combinations: P18 + DESI, P18 + (DESI+SDSS), P18 + DESI + SH0ES, and P18 + DESI + CCHP.}
\setlength{\tabcolsep}{8pt}
\begin{ruledtabular}
\begin{tabular} { l  c c c c}
\multicolumn{5}{c}{EMG-CC}\\
\midrule
{} & P18 + DESI & P18 + (DESI+SDSS) & P18 + DESI + SH0ES  & P18 + DESI + CCHP\\
\midrule
{\boldmath$\ln(10^{10} A_\mathrm{s})$}         & $3.052\pm 0.015       $ & $3.050\pm 0.014$             & $3.057\pm 0.015$                & $3.053\pm 0.015$\\
{\boldmath$n_\mathrm{s}$}                      & $0.9707\pm 0.0040     $ & $0.9686\pm 0.0038$           & $0.9758\pm 0.0040$              & $0.9715\pm 0.0039$\\
{\boldmath$H_0 [\mathrm{km\, s^{-1} Mpc^{-1}}]$} & $68.74^{+0.58}_{-0.72}$ & $68.24^{+0.47}_{-0.58}$      & $70.27\pm 0.65$                 & $68.95^{+0.60}_{-0.70}$\\
{\boldmath$\Omega_\mathrm{b} h^2$}             & $0.02248\pm 0.00014   $ & $0.02244\pm 0.00014$         & $0.02256^{+0.00013}_{-0.00015}$ & $0.02250\pm 0.00014$\\
{\boldmath$\Omega_\mathrm{c} h^2$}             & $0.1190\pm 0.0011     $ & $0.1195^{+0.0009}_{-0.0011}$ & $0.1191^{+0.0010}_{-0.0012}$    & $0.1189^{+0.0010}_{-0.0011}$\\
{\boldmath$\tau_\mathrm{reio}$}                & $0.0584\pm 0.0074     $ & $0.0569\pm 0.0072$           & $0.0594\pm 0.0076$              & $0.0586\pm 0.0075$\\
{\boldmath$\sigma_\mathrm{ini} [M_\mathrm{Pl}]$} & $ < 0.38$ (95\%)        & $< 0.34$ (95\%)              & $0.38^{+0.08}_{-0.06}$          & ${0.24^{+0.17}_{-0.23}}$ (95\%)\\
{\boldmath$V_0$}                               &  ---                    & ---                          & $< -1.2$ (95\%)                 & ---\\
\midrule
$\xi\sigma_\mathrm{ini}^2 [M_\mathrm{Pl}^2]$  & $> -0.024$ (95\%) & $> -0.019$ (95\%) & $-0.025\pm 0.009$ & $> -0.026$ (95\%) \\
$S_8$                                       & $0.819\pm 0.011$   & $0.824\pm 0.010$ & $0.816\pm 0.011$  & $0.818\pm 0.011$\\
\midrule
$\Delta\bigchi^2\mathrm{(P18)}$ & $-0.1$ & $-1.1$ & $-2.9$  & $-2.6$ \\ 
$\Delta\bigchi^2\mathrm{(BAO)}$ & $-1.8$ & $-0.8$ & $1.0$   & $-0.8$ \\ 
$\Delta\bigchi^2(H_0)$          & ---    & ---    & $-9.4$  & $-0.7$ \\ 
$\Delta\bigchi^2$               & $-2.0$ & $-1.9$ & $-11.4$ & $-4.1$ \\ 
$\Delta \rm AIC $               & $2.0 $ & $2.1$  & $-7.4$  & $-0.1$ \\ 
\end{tabular}
\end{ruledtabular}
\label{tab:emg_cc_table}
\end{table*} 

\clearpage

\section{Plots}\label{sec:plots}
\onecolumngrid
In this section we present the triangle plots of the 2D and 1D marginalized constraints on cosmological parameters, obtained in our analysis and with previous BAO datasets.

\begin{figure*}[b]
\label{fig:ig_tri_all_datasets}
\includegraphics[width=\textwidth]{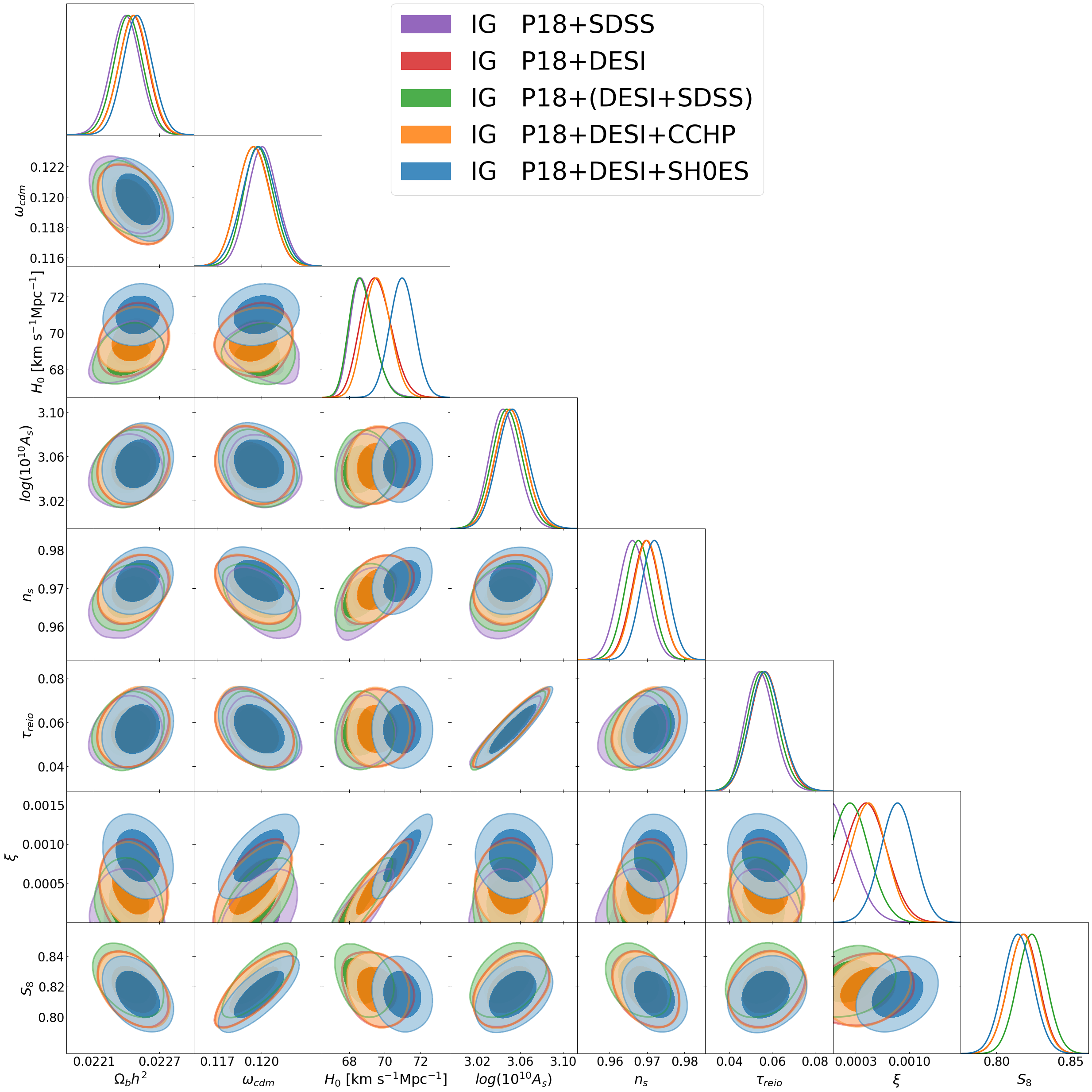}
\caption{Triangle plot in IG comparing results for P18+SDSS, P18+DESI, P18+(DESI+SDSS), P18+DESI+SH0ES, and P18+DESI+CCHP. 
The results for P18+SDSS are taken from \cite{Ballardini:2020iws}, where the SDSS dataset was composed of the BOSS DR12 \cite{BOSS:2016wmc} consensus results on BAOs in three redshift slices with effective redshifts $z_\mathrm{eff} = 0.38, 0.51, 0.61$ \cite{BOSS:2016apd,BOSS:2016sne,BOSS:2016hvq}, in combination with measure from 6dF \cite{2011MNRAS.416.3017B} at $z_\mathrm{eff} = 0.106$ and the one from SDSS DR7 \cite{Ross:2014qpa} at $z_\mathrm{eff} = 0.15$.}
\end{figure*}
 
\begin{figure*}
\label{fig:dig_tri_all_datasets}
\includegraphics[width=\textwidth]{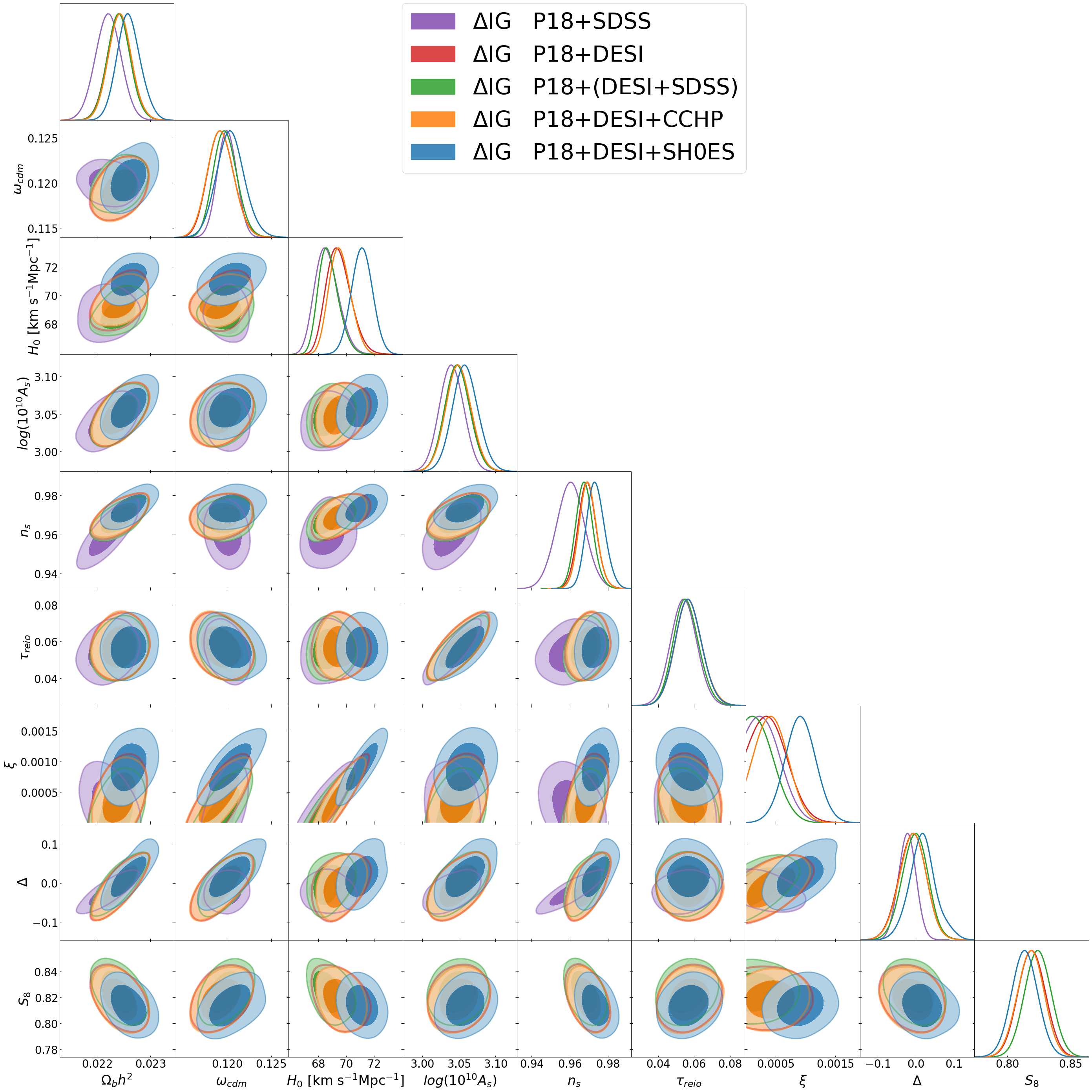}
\caption{Triangle plot in $\Delta$IG comparing results for P18+SDSS, P18+DESI, P18+(DESI+SDSS), P18+DESI+SH0ES, and P18+DESI+CCHP. 
The results for P18+SDSS are taken from \cite{Ballardini:2021evv}, where the SDSS dataset was composed of the BOSS DR12 \cite{BOSS:2016wmc} consensus results on BAOs in three redshift slices with effective redshifts $z_\mathrm{eff} = 0.38, 0.51, 0.61$ \cite{BOSS:2016apd,BOSS:2016sne,BOSS:2016hvq}, in combination with measure from 6dF \cite{2011MNRAS.416.3017B} at $z_\mathrm{eff} = 0.106$ and the one from SDSS DR7 \cite{Ross:2014qpa} at $z_\mathrm{eff} = 0.15$.}
\end{figure*}
 
\begin{figure*}
\label{fig:bdg_tri_all_datasets}
\includegraphics[width=\textwidth]{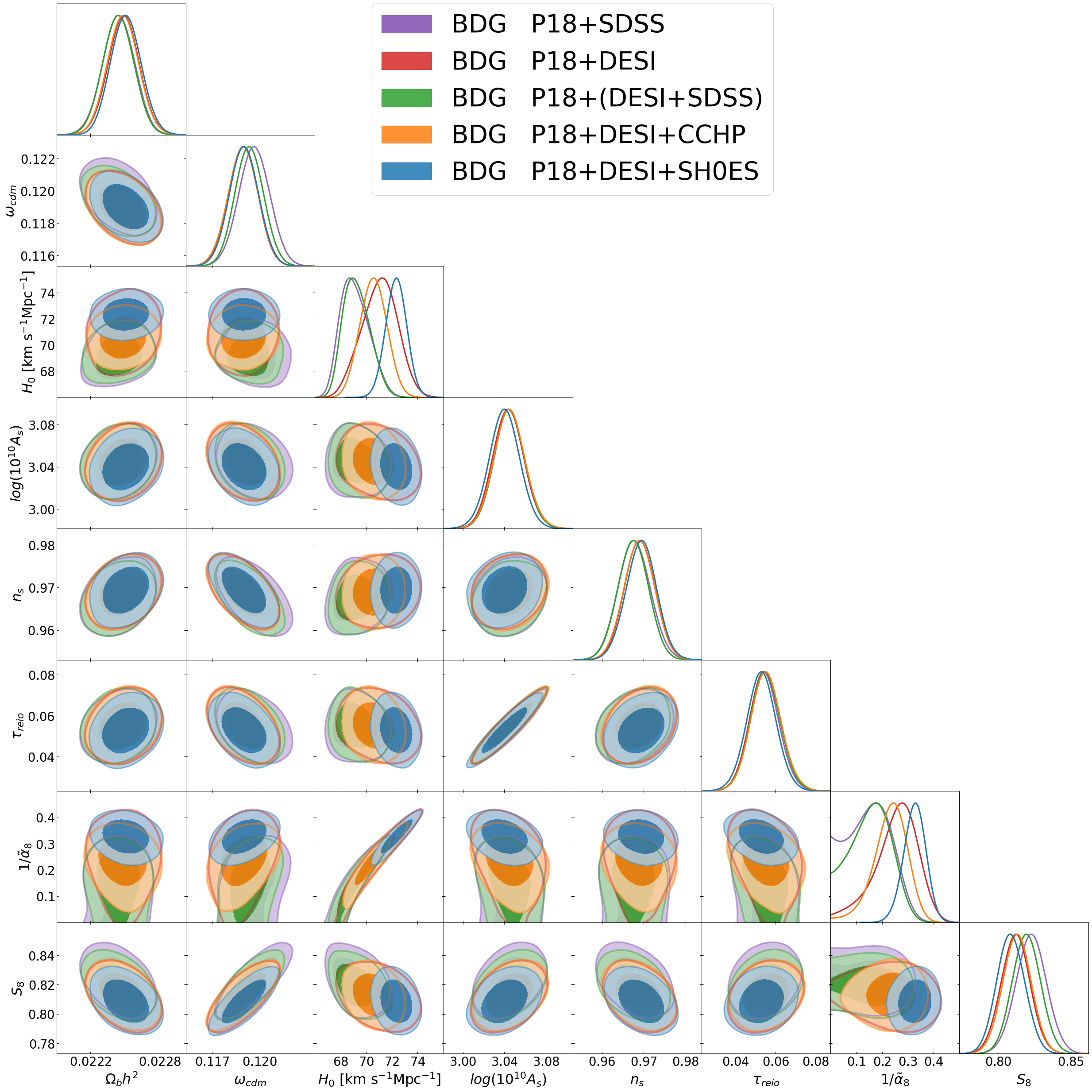}
\caption{Triangle plot in BDG comparing results for P18+SDSS, P18+DESI, P18+(DESI+SDSS), P18+DESI+SH0ES, and P18+DESI+CCHP. 
The results for P18+SDSS are taken from \cite{Ferrari:2023qnh}, where the SDSS dataset was composed of the postreconstruction measurements from BOSS DR12 \cite{BOSS:2016wmc}, low-$z$ BAO measurements from SDSS DR7, 6dF and MGS \cite{Ross:2014qpa,2011MNRAS.416.3017B}, Ly$\alpha$ BAO measurements from eBOSS DR14, and combination of those \cite{deSainteAgathe:2019voe,Blomqvist:2019rah,Cuceu:2019for}.}
\end{figure*}

\begin{figure*}
\label{fig:emg_cc_tri_all_datasets}
\includegraphics[width=\textwidth]{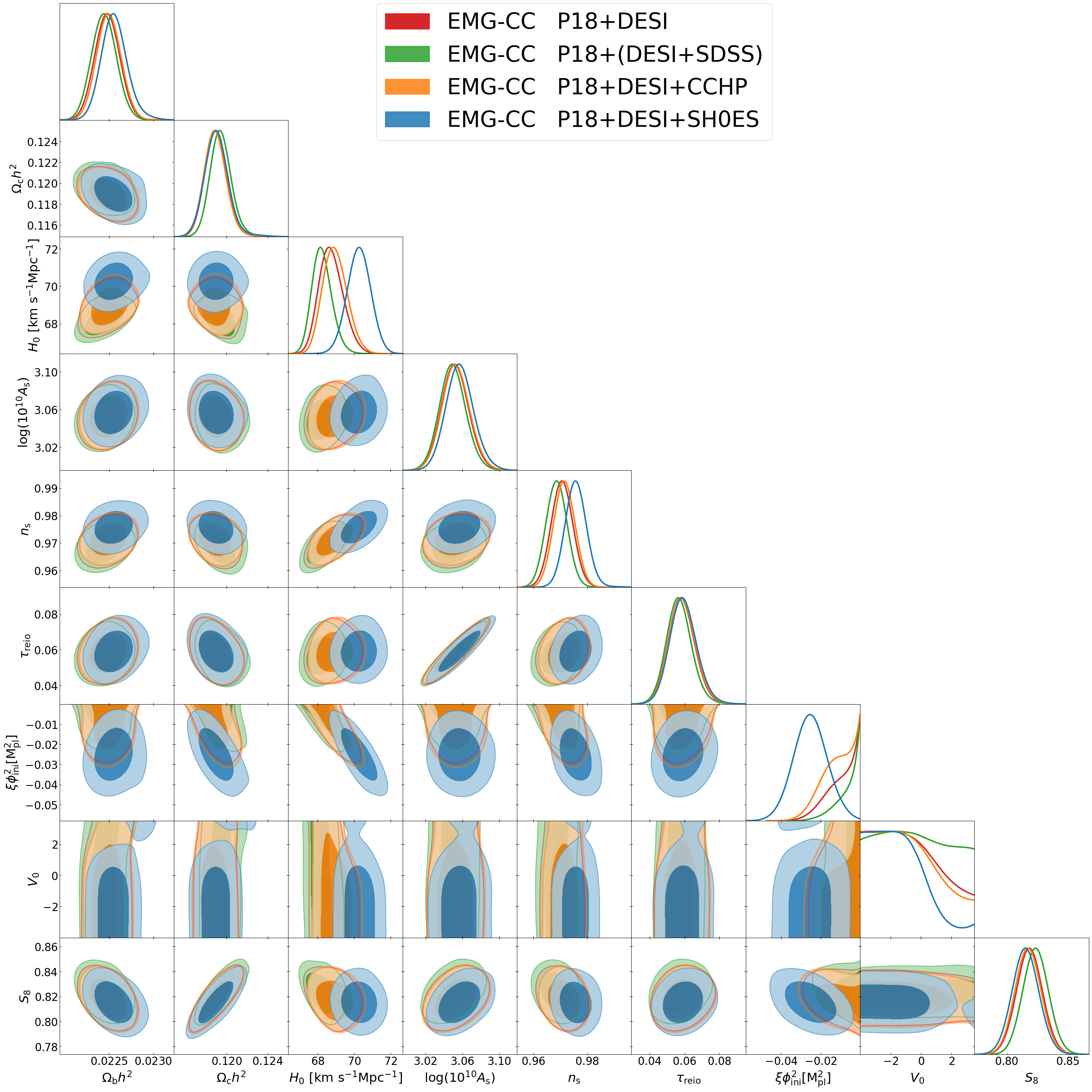}
\caption{Triangle plot in EMG-CC comparing results for P18+DESI, P18+(DESI+SDSS), P18+DESI+SH0ES, and P18+DESI+CCHP.}
\end{figure*}

\clearpage
\twocolumngrid
\section{Equations of motion in FLRW background}\label{sec:flrw_equations}

It is possible to write the equations that govern the background evolution by starting from \cref{eq:EE,eq:cova_KG} and specializing them to a FLRW metric, with line element
\begin{equation}
\dd s^2 = -\dd t^2 + a^2(t) \, \dd x^2 \,.
\end{equation}
In this setting, the covariant Einstein field equations \eqref{eq:EE} reduce to
\begin{align}
\begin{split}
3 F H^2 &= \rho + \frac{1}{2} Z \dot\sigma^2 - 3 H \dot{F} + V(\sigma) + \\
&+ \dot\sigma^3\left[6 g(\sigma) H - \dot{g}(\sigma)  + 3\zeta(\sigma) \dot\sigma \right] \equiv \rho + \rho_{\sigma}\,,
\end{split}
\label{eq:Friedmann}
\end{align}
\begin{align}
\begin{split}
-2 F \dot{H} = &\rho + p + Z\dot\sigma^2 + \ddot{F} - H\dot{F} \\
&+ \dot\sigma^2\left[(6 g H - 2 g,_\sigma\dot\sigma)\dot\sigma + 4\zeta\dot\sigma^2 -2 g \ddot\sigma\right]\\
\equiv &\rho + p + \rho_\sigma + p_\sigma \,,
\end{split}
\label{eq:Friedmann2}
\end{align}
where
\begin{equation}
\rho_\sigma = \frac{Z}{2} \dot\sigma^2 - 3 H \dot{F} + V(\sigma) 
		+ \dot\sigma^3\big[6 g(\sigma) H - \dot{g}(\sigma)  + 3\zeta(\sigma) \dot\sigma \big] \,,
\end{equation}
\begin{equation}
p_\sigma = \frac{Z}{2} \dot\sigma^2 - V(\sigma) + \ddot{F} + 2 H \dot{F} - \dot\sigma^4 ( g,_{\sigma} - \zeta ) - 2 g \dot\sigma^2 \ddot\sigma \,.
\end{equation}

The scalar field equation \eqref{eq:cova_KG} in the FLRW metric takes the following form
\begin{align}
\begin{split}
&\ddot\sigma \big(Z+12 g H \dot\sigma - 4 (g,_{\sigma} -3\zeta)\dot\sigma^2 \big)- 3F,_\sigma\,(2H^2+\dot{H}) \\
&+ V,_\sigma + 3 Z H \dot\sigma + 6 g (3H^2+\dot{H})\dot\sigma^2 +12 H \zeta \dot\sigma^3 \\
&- (g,_{\sigma\sigma} - 3\zeta,_{\sigma}) \dot\sigma^4 = 0 \,.
\label{eq:scalar_field_flrw}
\end{split}    
\end{align}

We define the density parameters for radiation (r), pressureless matter (m), and the scalar field ($\sigma$) following the notation of \cite{Boisseau:2000pr,Finelli:2007wb}
\begin{equation}
\widetilde\Omega_i = \frac{\rho_i}{3FH^2} \equiv \frac{\rho_i}{\rho_\mrm{crit}}\quad (i=\mrm{r, m}, \sigma) \,.
\label{eq:tilde_Omegas}
\end{equation}
It is also useful to define an effective dark-energy density and pressure parameters in a framework that mimics Einstein gravity at the present time: this is done by rewriting the Friedmann equations as \cite{Torres:2002pe,Gannouji:2006jm,Finelli:2007wb}
\begin{align}
3 F_0 H^2 &= \rho + \rho_\mrm{DE} \,,\\ 
-2 F_0 \dot{H} &= \rho  + p + \rho_\mrm{DE} + p_\mrm{DE} \,,
\end{align}
which leads to
\begin{align}
\rho_\mrm{DE} &= \frac{F_0}{F}\rho_{\sigma} + \rho\left(\frac{F_0}{F}-1\right) \,, \\
p_\mrm{DE} &= \frac{F_0}{F} p_{\sigma} + p\left(\frac{F_0}{F}-1\right) \,.
\end{align}
Thus, in this framework we can define the effective parameter of state for DE as 
\begin{equation}\label{eq:wde}
w_\mrm{DE}\equiv \frac{p_\mrm{DE}}{\rho_\mrm{DE}} \,,
\end{equation}
and the density parameters mimicking radiation, matter and DE in Einstein gravity are
\be
\Omega_i = \frac{\rho_i}{3 F_0 H^2} \quad (i=\mrm{r, m, DE}) \,.
\label{eq:Omegas}
\ee
Note that the definitions in \cref{eq:tilde_Omegas,eq:Omegas} coincide at the present time $z=0$: $\widetilde\Omega_{0, i}=\Omega_{0, i}$.
A plot of $w_\mathrm{DE}$ for the models considered is shown in \cref{fig:wde_all_models}.

Two quantities of interest are the variation of the cosmological gravitational constant between the radiation era and the present time 
\begin{equation}
\frac{\delta G_\mathrm{cosm}}{G_\mathrm{cosm}} (z=0) = \frac{G_\mathrm{cosm}(\sigma_0) - G_\mathrm{cosm} (\sigma_\mathrm{ini})}{G_\mathrm{cosm}(\sigma_\mathrm{ini})} = \frac{\sigma_\mathrm{ini}^2 - \sigma_0^2}{\sigma_0^2},
\end{equation}
and the time derivative of the gravitational constant at the present time
\begin{equation}
\frac{\dot{G}_\mathrm{cosm}}{G_\mathrm{cosm}} (z=0) = -\frac{2 \dot{\sigma}_0}{\sigma_0}.
\end{equation}

\clearpage
\bibliography{references}
\end{document}